\documentclass[aps,prl,reprint,superscriptaddress,twocolumn,showkeys,amsmath,amssymb,longbibliography,floatfix]{revtex4-2}
\usepackage[english]{babel}
\usepackage{amsmath,amssymb,bbm,mathrsfs,bm,braket,color,graphicx,comment,amsfonts,dsfont}
\usepackage[colorlinks,linkcolor=blue,citecolor=blue,urlcolor=blue]{hyperref}
\usepackage{hyperref}
\usepackage[mathscr]{euscript}
\usepackage{bm}

\begin{document}

\title{Two-point spectroscopy of Fibonacci topoelectrical circuits}

\author{Selma Franca}
\email{selma.franca@neel.cnrs.fr}
\affiliation{Institute for Theoretical Solid State Physics, IFW Dresden and W\"urzburg-Dresden Cluster of Excellence ct.qmat, Helmholtzstr. 20, 01069 Dresden, Germany}
\affiliation{Institut Neel, CNRS, Grenoble, France}

\author{Torsten Seidemann}
\affiliation{Institute for Theoretical Solid State Physics, IFW Dresden and W\"urzburg-Dresden Cluster of Excellence ct.qmat, Helmholtzstr. 20, 01069 Dresden, Germany}

\author{Fabian Hassler}
\affiliation{JARA-Institute for Quantum Information, RWTH Aachen University, 52056 Aachen, Germany}

\author{Jeroen van den Brink}
\affiliation{Institute for Theoretical Solid State Physics, IFW Dresden and W\"urzburg-Dresden Cluster of Excellence ct.qmat, Helmholtzstr. 20, 01069 Dresden, Germany}
\affiliation{Institute  for  Theoretical  Physics,  TU  Dresden,  01069  Dresden,  Germany}

\author{Ion Cosma Fulga}
\affiliation{Institute for Theoretical Solid State Physics, IFW Dresden and W\"urzburg-Dresden Cluster of Excellence ct.qmat, Helmholtzstr. 20, 01069 Dresden, Germany}

\newcommand{\noteCF}[1]{{\color[rgb]{0.,0.,0.8}[CF: #1]}}
\newcommand{\noteSF}[1]{{\color[rgb]{0,0.8,0.}[SF: #1]}}
\newcommand{\noteFH}[1]{{\color[rgb]{0.8,0.,0.}[FH: #1]}}

\date{\today}
\begin{abstract}
Topoelectrical circuits are meta-material realizations of topological features of condensed matter systems. 
In this work, we discuss experimental methods that allow a fast and straightforward detection of the spectral features of these systems from the two-point impedance of the circuit.
This allows to deduce the full spectrum of a topoelectrical circuit consisting of $N$ sites from a single two-point measurement of the frequency resolved impedance. 
In contrast, the standard methods rely on $N^2$ measurements of admittance matrix elements with a subsequent diagonalization on a computer.
We experimentally test our approach by constructing a Fibonacci topoelectrical circuit. 
Although the spectrum of this chain is fractal, i.e., more complex than the spectra of periodic systems, our approach is successful in recovering its eigenvalues.
Our work promotes the topoelectrical circuits as an ideal platform to measure spectral properties of various (quasi)crystalline systems. 
\end{abstract}
\maketitle

\textit{\textcolor{blue}{Introduction}} ---
Experimental difficulties in realizing and detecting topological features of condensed-matter systems
have prompted the development of meta-materials --- classical systems designed to reproduce desired topological features.
The initial proposal~\cite{Haldane2008} involved photonic crystals where electromagnetic waves propagate unidirectionally along the boundary, thus forming the photonic analog of the integer quantum Hall effect (IQHE)~\cite{vonKlitzing1980}. 
In addition to photonic meta-materials~\cite{Lu2014, Zeuner2015, Weimann2017, Noh2018, Saei2018, Chen2019, Mittal2019,El_Hassan2019}, there are acoustic~\cite{ Torrent2012, Yang2015, Fan2016, Xue2019, Ni2019, Apigo2019, Ni2019a, Chen2021}, mechanical~\cite{Susstrunk2015, Bertoldi2017, Serra-Garcia2018}, microwave~\cite{Bellec2013, Hu2015, Anderson2016, Peterson2018, Yu2020, Ma2020} and electrical circuit~\cite{Jia2015, Albert2015, Lee2018, Tal2018, Li2018, Hofmann2019, Haenel2019,  Zhu2019, Helbig2020, Hofmann2020, Wang2020, Rafi-Ul-Islam2020, Dong2021, Kotwal2021, Yang2023, Zheng2022, Zhang2023} realizations of various topological phases of matter.

Topoelectrical circuits are networks of nodes connected by electronic components such as resistors, capacitors, and inductors. 
They are described by an admittance matrix $Y (f)$ that represents the current response to a set $\textbf{V} (f)$ of locally applied voltages at frequency $f$, and that can be mapped to an effective tight-binding Hamiltonian~\cite{Lee2018, Imhof2018, Helbig2020}. 
So far, the experimental characterization of these classical systems mostly relied on detecting topological boundary phenomena using two-point impedance measurements~\cite{Lee2018, Imhof2018}.
This impedance, $Z_{a,b}(f)$, can be determined by measuring the voltage response between the nodes $a$ and $b$ to an input current oscillating at a specific frequency.
If this frequency corresponds to the energy of a topological boundary state of the effective Hamiltonian, and if the nodes are chosen such that one is in the bulk and the other in the region where this topological state is localized, the resulting two-point impedance is very large in realistic systems (it even diverges for ideal ones).
Thus, the presence of a topological boundary state inside of the bulk gap results in a single, isolated impedance peak.

Gaining access to the full spectrum of the effective Hamiltonian simulated by a topoelectrical circuit, beyond the detection of individual, spectrally isolated modes, is challenging.
The spectra of topoelectrical circuits have so far been determined by measuring the full admittance matrix, element by element, and then diagonalizing it on a computer~\cite{Stegmaier2021}.
This is a time-consuming process, since the number of measurements scales quadratically with the number of sites in the system, meaning that $N^2$ separate measurements are required for a circuit simulating a system made of $N$ sites.
Such disadvantageous scaling hinders the full spectrum measurement of a topoelectrical circuit, and undermines interest in realizing systems with intriguing spectral properties, like quasicrystals.

Quasicrystals are systems with incommensurate energy scales~\cite{Lenz2003, Pikovsky1995}, whose spectra may be fractal, resulting in local power law singularities of the associated density of states~\cite{Jagannathan2021}.
Since they are much rarer in nature, their meta-material realizations are even more relevant for studying their spectral properties~\cite{Jagannathan2021, Tanese2014}.
The prototypical example in one-dimension is the Fibonacci chain, an array of sites related by two possible hopping strengths arranged into a quasiperiodic pattern~\cite{Jagannathan2021}. 
Beyond its fractal spectrum, this chain is interesting because it can be adiabatically related to a two-dimensional Hofstadter model that realizes the IQHE physics. 
Consequently, the Fibonacci chain can support topological boundary states~\cite{Kraus2012, Verbin2013}.

In this work, we discuss a method that allows detection of an extensive number of topoelectrical circuit modes in two-point setup with fixed nodes. 
This method relies on measuring the linear response function of the circuit to a frequency-dependent input current.
We identify the eigenvalues of the effective tight-binding model~\cite{Stegmaier2021, Stegmaier2023, Zou2023} by determining the resonances of the two-point impedance through appropriate signal processing techniques. 
We test our approach under realistic conditions by constructing a topoelectrical Fibonacci chain. 
Despite having a fractal spectrum that is more complex than the spectrum of a periodic system,  we correctly identify most of the Fibonacci chain eigenvalues  in a single frequency resolved measurement by utilizing the spectral symmetry constraint imposed by the Fibonacci Hamiltonian.  

We start by introducing the Hamiltonian of the topoelectrical Fibonacci chain and showing how the linear response function is able to detect the eigenvalues of the circuit. 
We proceed with the experimental setup and discuss the measured data and corresponding numerical tools used to recover the Fibonacci chain spectrum.

\textit{\textcolor{blue}{Topoelectrical Fibonacci chain}} ---
In this work, we realize the $8$th approximant of the infinite quasiperiodic Fibonacci chain consisting of $N=34$ sites~\cite{Jagannathan2021}.
The Hamiltonian of the off-diagonal Fibonacci chain model is given by
\begin{equation}\label{eq:ham}
H(\phi) = \sum_{n=1}^N t_n (\phi) c_{n+1}^{\dagger} c_n^{} + \rm{h.c.},
\end{equation}
where $c_n^{\dagger}$ and $c_n^{}$ represent the creation and annihilation operator of a particle at site $n$.
The hoppings $t_n(\phi)= \alpha + \beta  \mathop{\rm sign}[\chi_n (\phi)]$ $(\alpha, \beta \in \mathbb{R})$ alternate between two values $t_A $ and $t_B$ as a function of the index $n$, such that $\alpha = (t_A+t_B)/2$ and $\beta=(t_A-t_B)/2$.
The alternation pattern is determined by the characteristic function $\chi_n (\phi)= \cos(\frac{2 \pi n}{\tau}+\phi) - \cos(\frac{\pi}{\tau})$ 
with the golden ratio $\tau = \frac{1+\sqrt{5}}{2}$ and the phason angle $\phi \in [0,2\pi)$~\cite{Kraus2012}. 
Setting $\phi = \pi$ creates the Fibonacci chain with two pairs of edge states that belong to different topological gaps.
These pairs of edge states occur at opposite energies because the Hamiltonian obeys the chiral symmetry constraint $\mathcal{C} H(\phi) \mathcal{C}^{\dagger} =-H(\phi)$ with $\mathcal{C}_{nm} = \delta_{nm} (-1)^{n}$.
Besides being symmetric with respect to zero energy, the Fibonacci chain spectrum is fractal~\cite{Suetoe1989}.
The eigenvalues are arranged in a self-similar pattern, as we can divide the spectrum into three clusters (or bands) of eigenvalues, and each cluster can be further split into three sub-clusters, and so on~\cite{Jagannathan2021}.

\begin{figure}[tb!]
\includegraphics[width=\columnwidth]{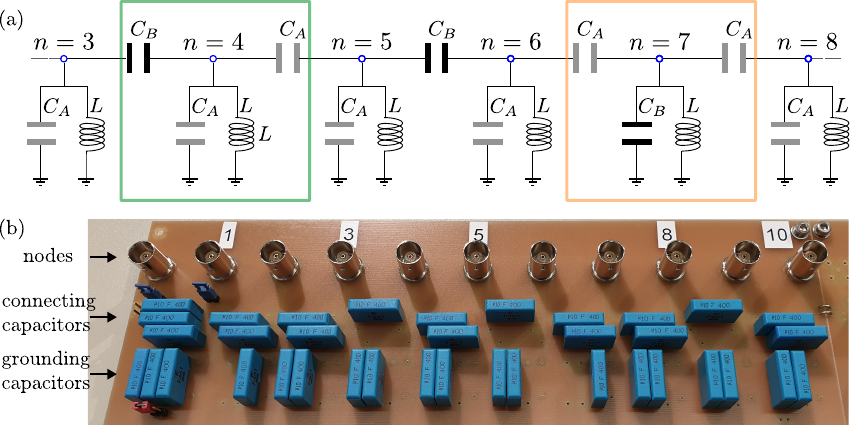}
\caption{Fibonacci topoelectrical circuit. 
(a) The circuit diagram between nodes $n=3$ and $n=8$. 
Orange and green boxes indicate two different configurations of topoloelectrical circuit junctions. 
(b) A photograph of the corresponding segment of the circuit board. 
We see all elements of the circuit diagram of panel (a), except inductors that are located on the backside. }
\label{fig:exp_nodes}
\end{figure}

In the following, we describe an electrical circuit that realizes the Fibonacci chain.
This circuit consists of $N=34$ nodes related by connecting wires and capacitors of distinct capacitances $C_A$ and $C_B$ that emulate the hoppings $t_A, t_B$ of the tight-binding model and are thus arranged according to $\mathop{\rm sign}[\chi_n(\pi)]$.
We show the circuit diagram inside the bulk of the system in Fig.~\ref{fig:exp_nodes}(a), and in Fig.~\ref{fig:exp_nodes}(b) the corresponding segment of a constructed circuit board.
The orange and green boxes in Fig.~\ref{fig:exp_nodes}(a) represent two possible local environments of bulk circuit nodes that differ by whether identical capacitances $(C_A, C_A)$ or distinct ones $(C_{A(B)}, C_{B(A)})$ are used to relate a node $n$ to its neighbors.
In the former (latter) case, for the grounding of node $n$ we use a capacitor of capacitance $\tilde{C}_n = C_B$ ($\tilde{C}_n=C_A$) that is connected in parallel to an inductor of inductance $L$, such that the relation $\tilde{C}_n + C_{n-1}+C_n = 2C_A+C_B$ holds.

Each node is described by Kirchhoff's law~\cite{Lee2018}
\begin{equation}\label{eq:klaw}
I_n = G_{n-1} (V_n-V_{n-1}) + G_n (V_n-V_{n+1})+ g_n V_n,
\end{equation}
where $G_{n} = 2\pi j f C_{n}$ is the admittance between nodes $n$ and $n+1$, $f$ is the frequency, $C_{n} \in \{C_A, C_B\}$ depending on $\mathop{\rm sign}[\chi_n(\pi)]$, and $j^2 = -1$.
The admittance $g_n$ between node $n$ and the ground equals $g_n = 2\pi j f \tilde{C}_{n}+1/(2\pi j f L)$, where conductance $\tilde{C}_{n} \in \{C_A, C_B\}$.
By grouping all currents and voltages into vectors $\mathbf{I}$ and $\mathbf{V}$, we obtain the admittance matrix $Y(f)$ 
\begin{equation} \label{eq:YtoH}
Y(f) = \tilde{g} (f) \mathbb{I} -2\pi j f H
\end{equation}
in terms of which Kirchhoff's rules are given by $\mathbf{I}(f) = Y(f) \mathbf{V}(f)$; here, $H$ is the Fibonacci Hamiltonian Eq.~\eqref{eq:ham} with hoppings $t_n$ replaced by $C_n$ and $\tilde{g}(f) = 2\pi j f (2C_A+C_B) +1/(2\pi j f L)$.

To experimentally characterize the spectral properties of this circuit, we measure its response to the applied current $I(f)$. The voltage at node $b$ is related to an input current at node $a$ via the two-point impedance
\begin{equation}\label{eq:imp_spect}
Z_{a,b} (f) = \frac{V_a (f)-V_b (f)}{I_a (f) } =  \sum_n \frac{|v_{n,a}-v_{n,b}|^2}{Y_n},
\end{equation}
that can be calculated from the eigenvalues $Y_n  (f) $ and eigenvectors $v_n  (f) $ of the admittance matrix~\cite{Lee2018}.

Next, we describe how $Z_{a,b} (f)$ can be used to reconstruct the Fibonacci chain spectrum. 
From Eq.~\eqref{eq:imp_spect}, we see that $Z_{a,b} (f)$ has a pole at frequency $f_n$ every time $Y_n (f_n) = 0$.
Using Eq.~\eqref{eq:YtoH}, we can relate the admittance matrix eigenvalues $Y_n$ to Hamiltonian eigenvalues $E_n$ as $Y_n  (f)= \tilde{g} (f) -2\pi j f E_n$.
Therefore, $Y_n (f_n) = 0 \rightarrow E_n = \tilde{g}(f_n)/2\pi j f_n$, resulting in
\begin{equation}\label{eq:freq_from_E}
E_n =  2C_A+C_B - \frac{1}{4\pi^2 L f^2_n};
\end{equation}
we note in passing, that the energies are measured in units of capacitance. 
Due to the relation \eqref{eq:freq_from_E}, reconstructing $E_n$ relies on identifying the resonance frequencies $f_n$ of the response function $Z_{a,b} (f)$. 
In the following, we describe how this can be done in practice.

\textit{\textcolor{blue}{Experimental setup and measurement analysis}} ---
For our experimental realization, we have used capacitors with nominal values of capacitances $C_A = 50 \,\rm nF$ and $C_B = 100 \,\rm nF$, and inductors with nominal inductances $L = 10
\,\rm \mu H$.
The capacitors and inductors are high quality components bought from KEMET and WURTH Elektronik, respectively, that were pre-selected to vary less than $2\%$ from the corresponding nominal values of conductances and inductances. 
Importantly, these circuit elements have small but non-vanishing direct current resistances $R_C^\text{dc} \approx 25 \,\rm m\Omega$ and $R_L^\text{dc} \approx 85 \,\rm m \Omega$. In case of the inductors the resistance is frequency dependent and goes from $R_L^\text{ac}   \approx 105 \,\rm m\Omega $  (at $50\,\rm{kHz}$) to $R_L^\text{ac}  \approx 308 \,\rm m\Omega$ (at $250\,\rm{kHz}$). 
For more details, see the Supplemental Material (SM)~\cite{SM}.

\begin{figure}[t!]
\includegraphics[width=1\columnwidth]{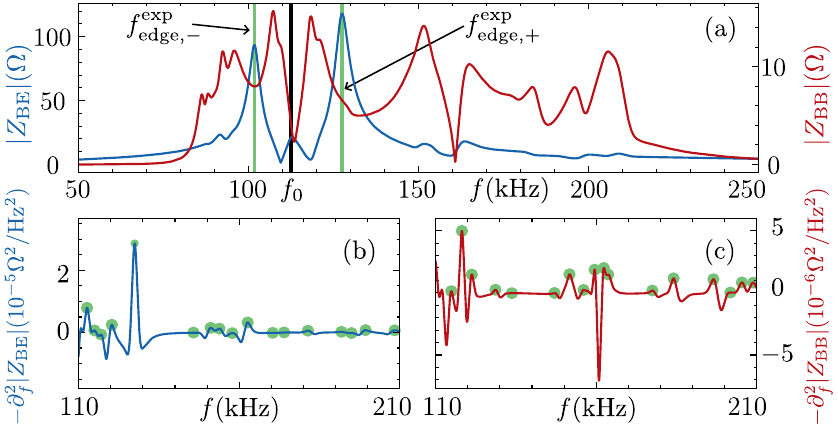}
\caption{Absolute values of the  measured impedances $|Z_{\rm BE}|$ and  $|Z_{\rm BB}|$  [panel (a)] and their second derivatives $-\partial_f^2 |Z_{\rm BE}|$ and $-\partial_f^2 |Z_{\rm BB}|$ [panels (b) and (c)] as a function of frequency $f$.  
The green circles indicate the peaks of $-\partial_f^2 |Z_{\rm BE}|$ and $-\partial_f^2 |Z_{\rm BB}|$ in the frequency range $(f_0, 250 \rm kHz)$, with $f_0 = 112.5 \rm kHz$.  }
\label{fig:impedance}
\end{figure}

All measurements were performed with the lock-in amplifier SR865A manufactured by Stanford Research Systems~\cite{SM}.
We consider two configurations for the voltage probes; the ``bulk-edge'' (BE) configuration is realized by placing probes at nodes $a=1$ and $b=15$, while the ``bulk-bulk'' (BB) configuration has the probes at nodes $a= 10$ and $b=24$. 
According to Eq.~\eqref{eq:imp_spect}, the positions of the voltage probes determine the weights of the corresponding eigenstates in the impedance response.
This results in a very different frequency dependence of the response functions $|{Z}_{\rm BE}|$ and $|{Z}_{\rm BB}|$ in range $f \in (50\rm \,kHz, 250\,\rm kHz)$, see Fig.~\ref{fig:impedance}(a).
To analyze these results, it is useful to define the frequency $f_0 =112.5\,$kHz corresponding to $E=0$ as determined from Eq.~\eqref{eq:freq_from_E} by setting $E_n = 0$ and using experimental values for $C_A, C_B$ and $L$.

Our first observation is that $|{Z}_{\rm BE}|$ and $|{Z}_{\rm BB}|$  have far fewer features for frequencies $f<f_0$ corresponding to energies $E<0$ than for $f>f_0$, where the positive part of the spectrum is located.
This is a consequence of the nonlinear relationship between the eigenvalues $E_n$ and resonant frequencies $f_n$ in Eq.~\eqref{eq:freq_from_E}. This leads to the fact that the resonant frequencies of the negative eigenvalues are closer together than the ones corresponding to the positive eigenvalues.
When this effect is combined with nonzero resistances $R_C^\text{dc}, R_L^\text{dc}$ and $R_L^{ac}$ that broaden the-delta-peaks of the ideal response function into Lorentzians, the resonant peaks for frequencies $f<f_0$ are expected to be less visible than the ones for $f>f_0$~\cite{SM}.
The second important feature of Fig.~\ref{fig:impedance}a is the observation that $|{Z}_{\rm BE}|$ has two very prominent peaks at frequencies (indicated by green lines) for which $|{Z}_{\rm BB}|$ does not show any prominent features.
This suggests that these peaks are induced by topological edge modes~\cite{Lee2018}.
From the corresponding frequencies $f^{\rm exp}_{\rm edge,-} \approx 101.7\,$kHz and $f^{\rm exp}_{\rm edge,+} \approx 127.5\,$kHz using Eq.~\eqref{eq:freq_from_E}, we obtain the energies $E^{\rm exp}_{\rm edge,-} = -44.9\,\rm nF$ and $E^{\rm exp}_{\rm edge,+} = 44.2\,\rm nF$.
Note that the theoretical value for energy of the edge states is $E^{}_{\rm{edge}, \pm} = \pm 43.7 \rm nF$; the relative errors are $\delta^r = |(E^{\rm exp}_{\rm edge, -}-E^{}_{\rm{edge}, -})/E^{}_{\rm{edge}, -}| = 2.75\%$ and $\delta^r = 1.14\%$, respectively.
Importantly, having $|E^{\rm exp}_{\rm edge,-}| \approx |E^{\rm exp}_{\rm edge,+}| \approx |E_{\rm edge, \pm}|$ is the experimental confirmation that the realized topoelectrical circuit has the chiral symmetry. 
We can use this symmetry to obtain an experimental value of the frequency $f_0^{\rm exp} = \frac12(f^{\rm exp}_{\rm edge,-}+f^{\rm exp}_{\rm edge,+}) =  114.6\,$kHz with a relative error $\delta^r = 1.87 \%$ compared to $f_0$.

\begin{figure*}[t!]
\includegraphics[width=1\textwidth]{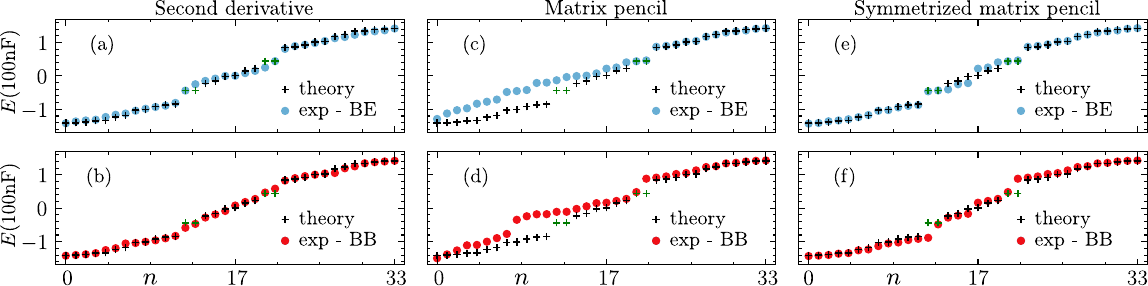}
\caption{
Comparison between the theoretical and experimental spectra obtained using different methods of recovery. 
Here, the green crosses indicate edge states.
The eigenvalues in panels (a) and (b) are given by the maxima of $-\partial_f^2 |Z_{\rm BE}|$ and $-\partial_f^2 |Z_{\rm BB}|$, see green circles in Figs.~\ref{fig:impedance}(b) and (c). 
In panels (c) and (d),  $34$ resonant frequencies are shown that were detected using the MP method. 
Panels (e) and (f) are obtained by mirroring the $17$ largest positive eigenvalues from panels (c) and (d) with respect to $E=0$.}
\label{fig:spectra}
\end{figure*}

To determine more eigenvalues, we focus on the second derivative of the response function because differentiation reduces the amplitude of broader peaks in the $Z_{a,b}(f)$ signal.
This results in a better detection of resonances that have been previously obscured by a broader but stronger background~\cite{OHaverBook}.
In practice, calculating this derivative from the original data set is challenging because measurements always include some noise that manifests as random high-frequency and small amplitude deviations from the ideal signal.
Since noise becomes more prominent with differentiation, we eliminate it from original data using a low-pass, $4$-th order Butterworth filter~\cite{Butterworth1930}.
This filter has a maximally flat frequency response in the passband, thus not giving rise to any additional frequency dependence upon its application~\cite{Butterworth1930}.

We employ two different strategies for extracting the Fibonacci chain spectrum.
Our first approach is based on searching for the frequencies $f^{\rm exp}_n$ at which the function $-\partial_f^2 |Z_{a,b}|$ (and consequently $|Z_{a,b}(f)|$) has peaks.
To calculate $-\partial_f^2 |Z_{a,b}|$, we employ the Butterworth filter with the cutoff frequency $f_c  = 0.01f_{\rm Nq}$ on $|Z_{a,b}(f)|$;
here, $f_{\rm Nq}$ denotes the Nyquist frequency defined as half of the sampling frequency $f$.
Due to aforementioned grouping effect of individual peaks in the lower frequency range, looking for $34$ most prominent peaks of $-\partial_f^2 |Z_{a,b}|$ in the entire frequency range does not produce satisfying results. 
Because of the chiral symmetry, we can instead focus on the frequency range $(f_0, 250\rm kHz)$ that corresponds to the positive part of the spectrum consisting of $17$ eigenvalues.   
Using the SCIPY Python library~\cite{Zenodo}, we find all the peaks of $-\partial_f^2 |Z_{\rm BE}|$ and $-\partial_f^2 |Z_{\rm BB}|$ in this frequency range and choose the $17$ most prominent ones for both curves. 
These peaks are indicated with green circles in Figs.~\ref{fig:impedance}(b) and (c) for $-\partial_f^2 |Z_{\rm BE}|$ and $-\partial_f^2 |Z_{\rm BB}|$, respectively.

The corresponding spectra are constructed from pairs $(-E^{\rm exp}_n, E^{\exp}_n)$, with $E^{\exp}_n$ obtained from $f^{\rm exp}_n$ using Eq.~\eqref{eq:freq_from_E}. 
We plot these spectra in Figs.~\ref{fig:spectra}(a) and (b) for the BE and BB voltage probe configurations, respectively, along with the theoretical eigenvalues $E_n$.
We observe that both voltage probes are successful in detecting the edges of the upper band (and consequently the lower band), along with its inner sub-bands.
The BE probe captures accurately the energies of two pairs of edge states but it detects a single resonance per pair. 
This behavior is also present for an ideal circuit, and originates from the energy degeneracy of two edge modes.
On the other side, the BB probe detects two resonances inside the topological gap but is less accurate in measuring the energies of edge states. 
In total, for the BE probe the mean absolute error $\delta^{\rm avg} = \sum_{n=1}^{N} |E_n-E_n^{\rm exp}|/N$ equals $\delta^{\rm avg}_{\rm BE} =4.87 \,\rm nF$ while the median error is $\delta_{\rm BE}^m = 3.81 \,\rm nF$. 
For the BB probe, we find $\delta^{\rm avg}_{\rm BB} = 4.17 \,\rm nF$ and $\delta_{\rm BB}^m = 3.18 \,\rm nF$.

As these errors are small compared to the total energy range, we conclude that searching for peaks of $-\partial_f^2 |Z_{a,b}|$ is a fruitful strategy to recover the full spectrum.
However, this approach focuses only on the amplitude of the frequency dependent response function, and thus misses possible information hidden in the corresponding phase component.  
To rectify this, we employ our second approach that is based on fitting the full signal $-\partial_f^2 Z_{a.b}$ to the linear combination of Lorentzians.
To eliminate noise from the data, we use the Butterworth filter separately on $\Re[Z_{a,b} (f)]$ and  $\Im[Z_{a,b}(f)]$ before calculating their second derivatives with respect to frequency and combining them to obtain $-\partial_f^2 Z_{a,b}$. 
We use frequency cutoffs $f_c = 0.03f_{\rm Nq}$ $(f_c =0.01 f_{\rm Nq})$ for the BE (BB) configuration of voltage probes. 
The resulting signal $-\partial_f^2 Z_{a,b}$ is Fourier transformed into the time domain signal $Z^{(2)}_{a,b} (t) =\mathcal{F} [ -\partial_f^2 Z_{a,b} ] $ that is fitted to a sum of $N$ damped exponentials as
\begin{equation}
Z^{(2)}_{a,b} (t) = \sum_{n=1}^N A^{\rm exp}_n  e^{j \phi^{\rm exp}_n} e^{(\alpha^{\rm exp}_n + 2\pi j f^{\rm exp}_n) t} ,
\end{equation}
where $A^{\rm exp}_n, \phi^{\rm exp}_n, \alpha^{\rm exp}_n$ and $f^{\rm exp}_n$ are the amplitudes, phases, damping factors and frequencies of the sinusoids, respectively.
Assuming $ t= m T$ where $m = 0,...,N-1$ and $T$ is the sampling period, the exponential factor becomes $e^{(\alpha^{\rm exp}_n + 2\pi j f^{\rm exp}_n) m T}= z_n^m$, where $z_n =e^{(\alpha^{\rm exp}_n + 2\pi j f^{\rm exp}_n) T} $. 
The poles $z_n$ are found by solving a generalized eigenvalue equation using a matrix pencil (MP) operator that is constructed from the values $Z^{(2)}_{a,b} (t)$~\cite{Hua1990, Vanhamme2001, Zieliski2011}, see also the SM~\cite{SM}. 
As there are $34$ eigenvalues in theory, we look for $34$ poles in our calculation.

The resulting spectrum, obtained using Eq.~\eqref{eq:freq_from_E}, is shown in Figs.~\ref{fig:spectra}(c) and (d) for both voltage probe configurations.
While this approach can reconstruct the entire spectrum, we see that for both configurations it works better for positive eigenvalues, i.e., for frequencies $f>f_0$. 
In general, the accuracy of the MP method declines as the energy is reduced which can be expected due to the aforementioned grouping effect of resonances. 
Moreover, for both probes, the method finds $15$ ($19$) poles corresponding to negative (positive) energies.
The additional positive poles arise at $E \sim 1\rm nF$ that is very close to $E=0$ in comparison with the energy scale of the chain.
In case of the BE probe where the edge modes dominate the response of the circuit, the MP method overestimates the number of edge modes in the upper topological gap but captures  their energies well. 
For the BB probe, the method finds a single mode for a pair of edge modes with $E>0$, and attributes the missing edge mode resonance to the upper band.
In total, we find $\delta^{\rm avg}_{\rm BE} = 21.67 \,\rm nF, \delta^{m}_{\rm BE} = 14.09 \,\rm nF$, and $\delta^{\rm avg}_{\rm BB} = 19.83 \,\rm nF, \delta^{m}_{\rm BB} = 11.54 \,\rm nF$.
Such large values of errors reflect the fact that the MP method misses to capture the negative eigenvalues accurately. 

The results of the MP-method can be improved by utilizing the spectral symmetry constraint, i.e., by constructing spectra from pairs $(-E^{\rm exp}_n, E^{\rm exp}_n)$, where $E^{\rm exp}_n>0$ are $17$ largest positive eigenvalues from Figs.~\ref{fig:spectra}(c) and (d).
Results are shown in Figs.~\ref{fig:spectra}(e) and (f) for BE and BB configurations, respectively.
This combined approach reduces the errors of measurements to $\delta^{\rm avg}_{\rm BE} = 4.24 \,\rm nF$, $\delta^{m}_{\rm BE} = 2.22 \,\rm nF$ for the BE probe and $\delta^{\rm avg}_{\rm BB} = 7.68 \,\rm nF$, $\delta^{m}_{\rm BB} = 5.22\, \rm nF$ for the BB probe.
Therefore, combining the MP method with the spectral symmetry constraint works the best for the BE probe, while searching for the peaks of $-\partial_f^2 |Z_{a,b}|$ yields better results for the BB probe.

For both probes, our best results have $\delta_{a,b}^{\rm avg} \approx \delta_E /2$, where $\delta_E = 8.55 \rm nF$ is a theoretical average energy spacing.
These results could be improved by reducing the noise in the measurement and the resistances of circuit elements.
Contrary to the present study that separately measured $V_a(f), V_b(f), I_a(f)$ thus increasing the chance of random events, employing additional lock-in amplifiers would allow for a simultaneous measurement of all three quantities and thus reduce the noise.
Reducing resistances of circuit elements, on the other hand, is not straightforward:
for example, lowering $R_L^\text{ac} (100\rm kHz)$ generally requires reducing the inductance $L$ of inductors, resulting in a larger frequency range needed to determine all the eigenvalues which leads to an increased of $R_L^\text{ac}$. 
As a result, the inductors will produce additional heating which washes out features due to the increased noise.
An interesting idea for future research is to investigate whether superconducting elements with significantly smaller resistances can improve the accuracy of our results.

\textit{\textcolor{blue}{Conclusion}} ---
In this work, we have shown how the response function of an electrical circuit can be used to recover the full spectrum of an underlying condensed matter system simulated by this circuit.
We have constructed, for the first time, the Fibonacci topoelectrical chain that has a fractal spectrum due to its quasicrystalline nature. 
We have demonstrated that the spectrum can be recovered from a single measurement using two distinct methods of data analysis.
We have corroborated our findings by changing the positions of the voltage probes as well as the boundary conditions of the Fibonacci topoelectrical circuit~\cite{SM}.
In conclusion, our work promotes topoelectrical circuits as an ideal meta-material platform for studying spectral properties of (quasi)crystalline systems. 

\textit{\textcolor{blue}{Acknowledgements}} ---
We thank Ulrike Nitzsche for technical assistance. 
This work was supported by the Deutsche Forschungsgemeinschaft (DFG, German Research Foundation) under Germany's Excellence Strategy through the W\"{u}rzburg-Dresden Cluster of Excellence on Complexity and Topology in Quantum Matter -- \emph{ct.qmat} (EXC 2147, project-id 390858490) and under Germany's Excellence Strategy -- Cluster of Excellence Matter and Light for Quantum Computing (ML4Q) EXC 2004/1 -- 390534769.
S. F. acknowledges financial support from the European Union Horizon 2020 research and innovation program under grant agreement No. 829044 (SCHINES).

\textit{\textcolor{blue}{Competing Interests Statement}} ---
The authors declare no competing interests.

\bibliography{NHr}

\begin{thebibliography}{62}%
\makeatletter
\providecommand \@ifxundefined [1]{%
 \@ifx{#1\undefined}
}%
\providecommand \@ifnum [1]{%
 \ifnum #1\expandafter \@firstoftwo
 \else \expandafter \@secondoftwo
 \fi
}%
\providecommand \@ifx [1]{%
 \ifx #1\expandafter \@firstoftwo
 \else \expandafter \@secondoftwo
 \fi
}%
\providecommand \natexlab [1]{#1}%
\providecommand \enquote  [1]{``#1''}%
\providecommand \bibnamefont  [1]{#1}%
\providecommand \bibfnamefont [1]{#1}%
\providecommand \citenamefont [1]{#1}%
\providecommand \href@noop [0]{\@secondoftwo}%
\providecommand \href [0]{\begingroup \@sanitize@url \@href}%
\providecommand \@href[1]{\@@startlink{#1}\@@href}%
\providecommand \@@href[1]{\endgroup#1\@@endlink}%
\providecommand \@sanitize@url [0]{\catcode `\\12\catcode `\$12\catcode
  `\&12\catcode `\#12\catcode `\^12\catcode `\_12\catcode `\%12\relax}%
\providecommand \@@startlink[1]{}%
\providecommand \@@endlink[0]{}%
\providecommand \url  [0]{\begingroup\@sanitize@url \@url }%
\providecommand \@url [1]{\endgroup\@href {#1}{\urlprefix }}%
\providecommand \urlprefix  [0]{URL }%
\providecommand \Eprint [0]{\href }%
\providecommand \doibase [0]{https://doi.org/}%
\providecommand \selectlanguage [0]{\@gobble}%
\providecommand \bibinfo  [0]{\@secondoftwo}%
\providecommand \bibfield  [0]{\@secondoftwo}%
\providecommand \translation [1]{[#1]}%
\providecommand \BibitemOpen [0]{}%
\providecommand \bibitemStop [0]{}%
\providecommand \bibitemNoStop [0]{.\EOS\space}%
\providecommand \EOS [0]{\spacefactor3000\relax}%
\providecommand \BibitemShut  [1]{\csname bibitem#1\endcsname}%
\let\auto@bib@innerbib\@empty
\bibitem [{\citenamefont {Haldane}\ and\ \citenamefont
  {Raghu}(2008)}]{Haldane2008}%
  \BibitemOpen
  \bibfield  {author} {\bibinfo {author} {\bibfnamefont {F.~D.~M.}\
  \bibnamefont {Haldane}}\ and\ \bibinfo {author} {\bibfnamefont
  {S.}~\bibnamefont {Raghu}},\ }\bibfield  {title} {\bibinfo {title} {Possible
  realization of directional optical waveguides in photonic crystals with
  broken time-reversal symmetry},\ }\href
  {https://doi.org/10.1103/PhysRevLett.100.013904} {\bibfield  {journal}
  {\bibinfo  {journal} {Phys. Rev. Lett.}\ }\textbf {\bibinfo {volume} {100}},\
  \bibinfo {pages} {013904} (\bibinfo {year} {2008})}\BibitemShut {NoStop}%
\bibitem [{\citenamefont {von Klitzing}\ \emph {et~al.}(1980)\citenamefont {von
  Klitzing}, \citenamefont {Dorda},\ and\ \citenamefont
  {Pepper}}]{vonKlitzing1980}%
  \BibitemOpen
  \bibfield  {author} {\bibinfo {author} {\bibfnamefont {K.}~\bibnamefont {von
  Klitzing}}, \bibinfo {author} {\bibfnamefont {G.}~\bibnamefont {Dorda}},\
  and\ \bibinfo {author} {\bibfnamefont {M.}~\bibnamefont {Pepper}},\
  }\bibfield  {title} {\bibinfo {title} {New method for high-accuracy
  determination of the fine-structure constant based on quantized {H}all
  resistance},\ }\href {https://doi.org/10.1103/PhysRevLett.45.494} {\bibfield
  {journal} {\bibinfo  {journal} {Phys. Rev. Lett.}\ }\textbf {\bibinfo
  {volume} {45}},\ \bibinfo {pages} {494} (\bibinfo {year} {1980})}\BibitemShut
  {NoStop}%
\bibitem [{\citenamefont {Lu}\ \emph {et~al.}(2014)\citenamefont {Lu},
  \citenamefont {Joannopoulos},\ and\ \citenamefont
  {Solja{\v{c}}i{\'{c}}}}]{Lu2014}%
  \BibitemOpen
  \bibfield  {author} {\bibinfo {author} {\bibfnamefont {L.}~\bibnamefont
  {Lu}}, \bibinfo {author} {\bibfnamefont {J.~D.}\ \bibnamefont
  {Joannopoulos}},\ and\ \bibinfo {author} {\bibfnamefont {M.}~\bibnamefont
  {Solja{\v{c}}i{\'{c}}}},\ }\bibfield  {title} {\bibinfo {title} {Topological
  photonics},\ }\href {https://doi.org/10.1038/nphoton.2014.248} {\bibfield
  {journal} {\bibinfo  {journal} {Nature Photon.}\ }\textbf {\bibinfo {volume}
  {8}},\ \bibinfo {pages} {821} (\bibinfo {year} {2014})}\BibitemShut {NoStop}%
\bibitem [{\citenamefont {Zeuner}\ \emph {et~al.}(2015)\citenamefont {Zeuner},
  \citenamefont {Rechtsman}, \citenamefont {Plotnik}, \citenamefont {Lumer},
  \citenamefont {Nolte}, \citenamefont {Rudner}, \citenamefont {Segev},\ and\
  \citenamefont {Szameit}}]{Zeuner2015}%
  \BibitemOpen
  \bibfield  {author} {\bibinfo {author} {\bibfnamefont {J.~M.}\ \bibnamefont
  {Zeuner}}, \bibinfo {author} {\bibfnamefont {M.~C.}\ \bibnamefont
  {Rechtsman}}, \bibinfo {author} {\bibfnamefont {Y.}~\bibnamefont {Plotnik}},
  \bibinfo {author} {\bibfnamefont {Y.}~\bibnamefont {Lumer}}, \bibinfo
  {author} {\bibfnamefont {S.}~\bibnamefont {Nolte}}, \bibinfo {author}
  {\bibfnamefont {M.~S.}\ \bibnamefont {Rudner}}, \bibinfo {author}
  {\bibfnamefont {M.}~\bibnamefont {Segev}},\ and\ \bibinfo {author}
  {\bibfnamefont {A.}~\bibnamefont {Szameit}},\ }\bibfield  {title} {\bibinfo
  {title} {Observation of a topological transition in the bulk of a
  non-{H}ermitian system},\ }\href
  {https://doi.org/10.1103/PhysRevLett.115.040402} {\bibfield  {journal}
  {\bibinfo  {journal} {Phys. Rev. Lett.}\ }\textbf {\bibinfo {volume} {115}},\
  \bibinfo {pages} {040402} (\bibinfo {year} {2015})}\BibitemShut {NoStop}%
\bibitem [{\citenamefont {Weimann}\ \emph {et~al.}(2017)\citenamefont
  {Weimann}, \citenamefont {Kremer}, \citenamefont {Plotnik}, \citenamefont
  {Lumer}, \citenamefont {Nolte}, \citenamefont {Makris}, \citenamefont
  {Segev}, \citenamefont {Rechtsman},\ and\ \citenamefont
  {Szameit}}]{Weimann2017}%
  \BibitemOpen
  \bibfield  {author} {\bibinfo {author} {\bibfnamefont {S.}~\bibnamefont
  {Weimann}}, \bibinfo {author} {\bibfnamefont {M.}~\bibnamefont {Kremer}},
  \bibinfo {author} {\bibfnamefont {Y.}~\bibnamefont {Plotnik}}, \bibinfo
  {author} {\bibfnamefont {Y.}~\bibnamefont {Lumer}}, \bibinfo {author}
  {\bibfnamefont {S.}~\bibnamefont {Nolte}}, \bibinfo {author} {\bibfnamefont
  {K.~G.}\ \bibnamefont {Makris}}, \bibinfo {author} {\bibfnamefont
  {M.}~\bibnamefont {Segev}}, \bibinfo {author} {\bibfnamefont {M.~C.}\
  \bibnamefont {Rechtsman}},\ and\ \bibinfo {author} {\bibfnamefont
  {A.}~\bibnamefont {Szameit}},\ }\bibfield  {title} {\bibinfo {title}
  {Topologically protected bound states in photonic parity--time-symmetric
  crystals},\ }\href {https://doi.org/10.1038/nmat4811} {\bibfield  {journal}
  {\bibinfo  {journal} {Nature Mater.}\ }\textbf {\bibinfo {volume} {16}},\
  \bibinfo {pages} {433} (\bibinfo {year} {2017})}\BibitemShut {NoStop}%
\bibitem [{\citenamefont {Noh}\ \emph {et~al.}(2018)\citenamefont {Noh},
  \citenamefont {Benalcazar}, \citenamefont {Huang}, \citenamefont {Collins},
  \citenamefont {Chen}, \citenamefont {Hughes},\ and\ \citenamefont
  {Rechtsman}}]{Noh2018}%
  \BibitemOpen
  \bibfield  {author} {\bibinfo {author} {\bibfnamefont {J.}~\bibnamefont
  {Noh}}, \bibinfo {author} {\bibfnamefont {W.~A.}\ \bibnamefont {Benalcazar}},
  \bibinfo {author} {\bibfnamefont {S.}~\bibnamefont {Huang}}, \bibinfo
  {author} {\bibfnamefont {M.~J.}\ \bibnamefont {Collins}}, \bibinfo {author}
  {\bibfnamefont {K.~P.}\ \bibnamefont {Chen}}, \bibinfo {author}
  {\bibfnamefont {T.~L.}\ \bibnamefont {Hughes}},\ and\ \bibinfo {author}
  {\bibfnamefont {M.~C.}\ \bibnamefont {Rechtsman}},\ }\bibfield  {title}
  {\bibinfo {title} {Topological protection of photonic mid-gap defect modes},\
  }\href {https://doi.org/10.1038/s41566-018-0179-3} {\bibfield  {journal}
  {\bibinfo  {journal} {Nature Photon.}\ }\textbf {\bibinfo {volume} {12}},\
  \bibinfo {pages} {408} (\bibinfo {year} {2018})}\BibitemShut {NoStop}%
\bibitem [{\citenamefont {Saei Ghareh~Naz}\ \emph {et~al.}(2018)\citenamefont
  {Saei Ghareh~Naz}, \citenamefont {Fulga}, \citenamefont {Ma}, \citenamefont
  {Schmidt},\ and\ \citenamefont {van~den Brink}}]{Saei2018}%
  \BibitemOpen
  \bibfield  {author} {\bibinfo {author} {\bibfnamefont {E.}~\bibnamefont {Saei
  Ghareh~Naz}}, \bibinfo {author} {\bibfnamefont {I.~C.}\ \bibnamefont
  {Fulga}}, \bibinfo {author} {\bibfnamefont {L.}~\bibnamefont {Ma}}, \bibinfo
  {author} {\bibfnamefont {O.~G.}\ \bibnamefont {Schmidt}},\ and\ \bibinfo
  {author} {\bibfnamefont {J.}~\bibnamefont {van~den Brink}},\ }\bibfield
  {title} {\bibinfo {title} {Topological phase transition in a stretchable
  photonic crystal},\ }\href {https://doi.org/10.1103/PhysRevA.98.033830}
  {\bibfield  {journal} {\bibinfo  {journal} {Phys. Rev. A}\ }\textbf {\bibinfo
  {volume} {98}},\ \bibinfo {pages} {033830} (\bibinfo {year}
  {2018})}\BibitemShut {NoStop}%
\bibitem [{\citenamefont {Chen}\ \emph {et~al.}(2019)\citenamefont {Chen},
  \citenamefont {Deng}, \citenamefont {Shi}, \citenamefont {Zhao},
  \citenamefont {Chen},\ and\ \citenamefont {Dong}}]{Chen2019}%
  \BibitemOpen
  \bibfield  {author} {\bibinfo {author} {\bibfnamefont {X.-D.}\ \bibnamefont
  {Chen}}, \bibinfo {author} {\bibfnamefont {W.-M.}\ \bibnamefont {Deng}},
  \bibinfo {author} {\bibfnamefont {F.-L.}\ \bibnamefont {Shi}}, \bibinfo
  {author} {\bibfnamefont {F.-L.}\ \bibnamefont {Zhao}}, \bibinfo {author}
  {\bibfnamefont {M.}~\bibnamefont {Chen}},\ and\ \bibinfo {author}
  {\bibfnamefont {J.-W.}\ \bibnamefont {Dong}},\ }\bibfield  {title} {\bibinfo
  {title} {Direct observation of corner states in second-order topological
  photonic crystal slabs},\ }\href
  {https://doi.org/10.1103/PhysRevLett.122.233902} {\bibfield  {journal}
  {\bibinfo  {journal} {Phys. Rev. Lett.}\ }\textbf {\bibinfo {volume} {122}},\
  \bibinfo {pages} {233902} (\bibinfo {year} {2019})}\BibitemShut {NoStop}%
\bibitem [{\citenamefont {Mittal}\ \emph {et~al.}(2019)\citenamefont {Mittal},
  \citenamefont {Orre}, \citenamefont {Zhu}, \citenamefont {Gorlach},
  \citenamefont {Poddubny},\ and\ \citenamefont {Hafezi}}]{Mittal2019}%
  \BibitemOpen
  \bibfield  {author} {\bibinfo {author} {\bibfnamefont {S.}~\bibnamefont
  {Mittal}}, \bibinfo {author} {\bibfnamefont {V.~V.}\ \bibnamefont {Orre}},
  \bibinfo {author} {\bibfnamefont {G.}~\bibnamefont {Zhu}}, \bibinfo {author}
  {\bibfnamefont {M.~A.}\ \bibnamefont {Gorlach}}, \bibinfo {author}
  {\bibfnamefont {A.}~\bibnamefont {Poddubny}},\ and\ \bibinfo {author}
  {\bibfnamefont {M.}~\bibnamefont {Hafezi}},\ }\bibfield  {title} {\bibinfo
  {title} {Photonic quadrupole topological phases},\ }\href
  {https://doi.org/https://doi.org/10.1038/s41566-019-0452-0} {\bibfield
  {journal} {\bibinfo  {journal} {Nature Photon.}\ }\textbf {\bibinfo {volume}
  {13}},\ \bibinfo {pages} {692} (\bibinfo {year} {2019})}\BibitemShut
  {NoStop}%
\bibitem [{\citenamefont {El~Hassan}\ \emph {et~al.}(2019)\citenamefont
  {El~Hassan}, \citenamefont {Kunst}, \citenamefont {Moritz}, \citenamefont
  {Andler}, \citenamefont {Bergholtz},\ and\ \citenamefont
  {Bourennane}}]{El_Hassan2019}%
  \BibitemOpen
  \bibfield  {author} {\bibinfo {author} {\bibfnamefont {A.}~\bibnamefont
  {El~Hassan}}, \bibinfo {author} {\bibfnamefont {F.~K.}\ \bibnamefont
  {Kunst}}, \bibinfo {author} {\bibfnamefont {A.}~\bibnamefont {Moritz}},
  \bibinfo {author} {\bibfnamefont {G.}~\bibnamefont {Andler}}, \bibinfo
  {author} {\bibfnamefont {E.~J.}\ \bibnamefont {Bergholtz}},\ and\ \bibinfo
  {author} {\bibfnamefont {M.}~\bibnamefont {Bourennane}},\ }\bibfield  {title}
  {\bibinfo {title} {Corner states of light in photonic waveguides},\ }\href
  {https://doi.org/10.1038/s41566-019-0519-y} {\bibfield  {journal} {\bibinfo
  {journal} {Nature Photon.}\ }\textbf {\bibinfo {volume} {13}},\ \bibinfo
  {pages} {697} (\bibinfo {year} {2019})}\BibitemShut {NoStop}%
\bibitem [{\citenamefont {Torrent}\ and\ \citenamefont
  {S\'anchez-Dehesa}(2012)}]{Torrent2012}%
  \BibitemOpen
  \bibfield  {author} {\bibinfo {author} {\bibfnamefont {D.}~\bibnamefont
  {Torrent}}\ and\ \bibinfo {author} {\bibfnamefont {J.}~\bibnamefont
  {S\'anchez-Dehesa}},\ }\bibfield  {title} {\bibinfo {title} {Acoustic
  analogue of graphene: Observation of {D}irac cones in acoustic surface
  waves},\ }\href {https://doi.org/10.1103/PhysRevLett.108.174301} {\bibfield
  {journal} {\bibinfo  {journal} {Phys. Rev. Lett.}\ }\textbf {\bibinfo
  {volume} {108}},\ \bibinfo {pages} {174301} (\bibinfo {year}
  {2012})}\BibitemShut {NoStop}%
\bibitem [{\citenamefont {Yang}\ \emph {et~al.}(2015)\citenamefont {Yang},
  \citenamefont {Gao}, \citenamefont {Shi}, \citenamefont {Lin}, \citenamefont
  {Gao}, \citenamefont {Chong},\ and\ \citenamefont {Zhang}}]{Yang2015}%
  \BibitemOpen
  \bibfield  {author} {\bibinfo {author} {\bibfnamefont {Z.}~\bibnamefont
  {Yang}}, \bibinfo {author} {\bibfnamefont {F.}~\bibnamefont {Gao}}, \bibinfo
  {author} {\bibfnamefont {X.}~\bibnamefont {Shi}}, \bibinfo {author}
  {\bibfnamefont {X.}~\bibnamefont {Lin}}, \bibinfo {author} {\bibfnamefont
  {Z.}~\bibnamefont {Gao}}, \bibinfo {author} {\bibfnamefont {Y.}~\bibnamefont
  {Chong}},\ and\ \bibinfo {author} {\bibfnamefont {B.}~\bibnamefont {Zhang}},\
  }\bibfield  {title} {\bibinfo {title} {Topological acoustics},\ }\href
  {https://doi.org/10.1103/PhysRevLett.114.114301} {\bibfield  {journal}
  {\bibinfo  {journal} {Phys. Rev. Lett.}\ }\textbf {\bibinfo {volume} {114}},\
  \bibinfo {pages} {114301} (\bibinfo {year} {2015})}\BibitemShut {NoStop}%
\bibitem [{\citenamefont {Fan}\ \emph {et~al.}(2016)\citenamefont {Fan},
  \citenamefont {Yu}, \citenamefont {Zhang}, \citenamefont {Zhang},\ and\
  \citenamefont {Ding}}]{Fan2016}%
  \BibitemOpen
  \bibfield  {author} {\bibinfo {author} {\bibfnamefont {L.}~\bibnamefont
  {Fan}}, \bibinfo {author} {\bibfnamefont {W.-w.}\ \bibnamefont {Yu}},
  \bibinfo {author} {\bibfnamefont {S.-y.}\ \bibnamefont {Zhang}}, \bibinfo
  {author} {\bibfnamefont {H.}~\bibnamefont {Zhang}},\ and\ \bibinfo {author}
  {\bibfnamefont {J.}~\bibnamefont {Ding}},\ }\bibfield  {title} {\bibinfo
  {title} {Zak phases and band properties in acoustic metamaterials with
  negative modulus or negative density},\ }\href
  {https://doi.org/10.1103/PhysRevB.94.174307} {\bibfield  {journal} {\bibinfo
  {journal} {Phys. Rev. B}\ }\textbf {\bibinfo {volume} {94}},\ \bibinfo
  {pages} {174307} (\bibinfo {year} {2016})}\BibitemShut {NoStop}%
\bibitem [{\citenamefont {Xue}\ \emph {et~al.}(2019)\citenamefont {Xue},
  \citenamefont {Yang}, \citenamefont {Gao}, \citenamefont {Chong},\ and\
  \citenamefont {Zhang}}]{Xue2019}%
  \BibitemOpen
  \bibfield  {author} {\bibinfo {author} {\bibfnamefont {H.}~\bibnamefont
  {Xue}}, \bibinfo {author} {\bibfnamefont {Y.}~\bibnamefont {Yang}}, \bibinfo
  {author} {\bibfnamefont {F.}~\bibnamefont {Gao}}, \bibinfo {author}
  {\bibfnamefont {Y.}~\bibnamefont {Chong}},\ and\ \bibinfo {author}
  {\bibfnamefont {B.}~\bibnamefont {Zhang}},\ }\bibfield  {title} {\bibinfo
  {title} {Acoustic higher-order topological insulator on a kagome lattice},\
  }\href {https://doi.org/10.1038/s41563-018-0251-x} {\bibfield  {journal}
  {\bibinfo  {journal} {Nature Mater.}\ }\textbf {\bibinfo {volume} {18}},\
  \bibinfo {pages} {108} (\bibinfo {year} {2019})}\BibitemShut {NoStop}%
\bibitem [{\citenamefont {Ni}\ \emph {et~al.}(2019{\natexlab{a}})\citenamefont
  {Ni}, \citenamefont {Weiner}, \citenamefont {Al{\`u}},\ and\ \citenamefont
  {Khanikaev}}]{Ni2019}%
  \BibitemOpen
  \bibfield  {author} {\bibinfo {author} {\bibfnamefont {X.}~\bibnamefont
  {Ni}}, \bibinfo {author} {\bibfnamefont {M.}~\bibnamefont {Weiner}}, \bibinfo
  {author} {\bibfnamefont {A.}~\bibnamefont {Al{\`u}}},\ and\ \bibinfo {author}
  {\bibfnamefont {A.~B.}\ \bibnamefont {Khanikaev}},\ }\bibfield  {title}
  {\bibinfo {title} {Observation of higher-order topological acoustic states
  protected by generalized chiral symmetry},\ }\href
  {https://doi.org/10.1038/s41563-018-0252-9} {\bibfield  {journal} {\bibinfo
  {journal} {Nature Mater.}\ }\textbf {\bibinfo {volume} {18}},\ \bibinfo
  {pages} {113} (\bibinfo {year} {2019}{\natexlab{a}})}\BibitemShut {NoStop}%
\bibitem [{\citenamefont {Apigo}\ \emph {et~al.}(2019)\citenamefont {Apigo},
  \citenamefont {Cheng}, \citenamefont {Dobiszewski}, \citenamefont {Prodan},\
  and\ \citenamefont {Prodan}}]{Apigo2019}%
  \BibitemOpen
  \bibfield  {author} {\bibinfo {author} {\bibfnamefont {D.~J.}\ \bibnamefont
  {Apigo}}, \bibinfo {author} {\bibfnamefont {W.}~\bibnamefont {Cheng}},
  \bibinfo {author} {\bibfnamefont {K.~F.}\ \bibnamefont {Dobiszewski}},
  \bibinfo {author} {\bibfnamefont {E.}~\bibnamefont {Prodan}},\ and\ \bibinfo
  {author} {\bibfnamefont {C.}~\bibnamefont {Prodan}},\ }\bibfield  {title}
  {\bibinfo {title} {Observation of topological edge modes in a quasiperiodic
  acoustic waveguide},\ }\href {https://doi.org/10.1103/PhysRevLett.122.095501}
  {\bibfield  {journal} {\bibinfo  {journal} {Phys. Rev. Lett.}\ }\textbf
  {\bibinfo {volume} {122}},\ \bibinfo {pages} {095501} (\bibinfo {year}
  {2019})}\BibitemShut {NoStop}%
\bibitem [{\citenamefont {Ni}\ \emph {et~al.}(2019{\natexlab{b}})\citenamefont
  {Ni}, \citenamefont {Chen}, \citenamefont {Weiner}, \citenamefont {Apigo},
  \citenamefont {Prodan}, \citenamefont {Al{\`u}}, \citenamefont {Prodan},\
  and\ \citenamefont {Khanikaev}}]{Ni2019a}%
  \BibitemOpen
  \bibfield  {author} {\bibinfo {author} {\bibfnamefont {X.}~\bibnamefont
  {Ni}}, \bibinfo {author} {\bibfnamefont {K.}~\bibnamefont {Chen}}, \bibinfo
  {author} {\bibfnamefont {M.}~\bibnamefont {Weiner}}, \bibinfo {author}
  {\bibfnamefont {D.~J.}\ \bibnamefont {Apigo}}, \bibinfo {author}
  {\bibfnamefont {C.}~\bibnamefont {Prodan}}, \bibinfo {author} {\bibfnamefont
  {A.}~\bibnamefont {Al{\`u}}}, \bibinfo {author} {\bibfnamefont
  {E.}~\bibnamefont {Prodan}},\ and\ \bibinfo {author} {\bibfnamefont {A.~B.}\
  \bibnamefont {Khanikaev}},\ }\bibfield  {title} {\bibinfo {title}
  {Observation of {H}ofstadter butterfly and topological edge states in
  reconfigurable quasi-periodic acoustic crystals},\ }\href
  {https://doi.org/10.1038/s42005-019-0151-7} {\bibfield  {journal} {\bibinfo
  {journal} {Commun. Phys.}\ }\textbf {\bibinfo {volume} {2}},\ \bibinfo
  {pages} {55} (\bibinfo {year} {2019}{\natexlab{b}})}\BibitemShut {NoStop}%
\bibitem [{\citenamefont {Chen}\ \emph {et~al.}(2021)\citenamefont {Chen},
  \citenamefont {Tang}, \citenamefont {Zhang}, \citenamefont {Chen},\ and\
  \citenamefont {Ma}}]{Chen2021}%
  \BibitemOpen
  \bibfield  {author} {\bibinfo {author} {\bibfnamefont {Z.-G.}\ \bibnamefont
  {Chen}}, \bibinfo {author} {\bibfnamefont {W.}~\bibnamefont {Tang}}, \bibinfo
  {author} {\bibfnamefont {R.-Y.}\ \bibnamefont {Zhang}}, \bibinfo {author}
  {\bibfnamefont {Z.}~\bibnamefont {Chen}},\ and\ \bibinfo {author}
  {\bibfnamefont {G.}~\bibnamefont {Ma}},\ }\bibfield  {title} {\bibinfo
  {title} {Landau-{Z}ener transition in the dynamic transfer of acoustic
  topological states},\ }\href {https://doi.org/10.1103/PhysRevLett.126.054301}
  {\bibfield  {journal} {\bibinfo  {journal} {Phys. Rev. Lett.}\ }\textbf
  {\bibinfo {volume} {126}},\ \bibinfo {pages} {054301} (\bibinfo {year}
  {2021})}\BibitemShut {NoStop}%
\bibitem [{\citenamefont {Süsstrunk}\ and\ \citenamefont
  {Huber}(2015)}]{Susstrunk2015}%
  \BibitemOpen
  \bibfield  {author} {\bibinfo {author} {\bibfnamefont {R.}~\bibnamefont
  {Süsstrunk}}\ and\ \bibinfo {author} {\bibfnamefont {S.~D.}\ \bibnamefont
  {Huber}},\ }\bibfield  {title} {\bibinfo {title} {Observation of phononic
  helical edge states in a mechanical topological insulator},\ }\href
  {https://doi.org/10.1126/science.aab0239} {\bibfield  {journal} {\bibinfo
  {journal} {Science}\ }\textbf {\bibinfo {volume} {349}},\ \bibinfo {pages}
  {47} (\bibinfo {year} {2015})}\BibitemShut {NoStop}%
\bibitem [{\citenamefont {Bertoldi}\ \emph {et~al.}(2017)\citenamefont
  {Bertoldi}, \citenamefont {Vitelli}, \citenamefont {Christensen},\ and\
  \citenamefont {van Hecke}}]{Bertoldi2017}%
  \BibitemOpen
  \bibfield  {author} {\bibinfo {author} {\bibfnamefont {K.}~\bibnamefont
  {Bertoldi}}, \bibinfo {author} {\bibfnamefont {V.}~\bibnamefont {Vitelli}},
  \bibinfo {author} {\bibfnamefont {J.}~\bibnamefont {Christensen}},\ and\
  \bibinfo {author} {\bibfnamefont {M.}~\bibnamefont {van Hecke}},\ }\bibfield
  {title} {\bibinfo {title} {Flexible mechanical metamaterials},\ }\href
  {https://doi.org/10.1038/natrevmats.2017.66} {\bibfield  {journal} {\bibinfo
  {journal} {Nat. Rev. Mater.}\ }\textbf {\bibinfo {volume} {2}},\ \bibinfo
  {pages} {17066} (\bibinfo {year} {2017})}\BibitemShut {NoStop}%
\bibitem [{\citenamefont {Serra-Garcia}\ \emph {et~al.}(2018)\citenamefont
  {Serra-Garcia}, \citenamefont {Peri}, \citenamefont {S{\"u}sstrunk},
  \citenamefont {Bilal}, \citenamefont {Larsen}, \citenamefont {Villanueva},\
  and\ \citenamefont {Huber}}]{Serra-Garcia2018}%
  \BibitemOpen
  \bibfield  {author} {\bibinfo {author} {\bibfnamefont {M.}~\bibnamefont
  {Serra-Garcia}}, \bibinfo {author} {\bibfnamefont {V.}~\bibnamefont {Peri}},
  \bibinfo {author} {\bibfnamefont {R.}~\bibnamefont {S{\"u}sstrunk}}, \bibinfo
  {author} {\bibfnamefont {O.~R.}\ \bibnamefont {Bilal}}, \bibinfo {author}
  {\bibfnamefont {T.}~\bibnamefont {Larsen}}, \bibinfo {author} {\bibfnamefont
  {L.~G.}\ \bibnamefont {Villanueva}},\ and\ \bibinfo {author} {\bibfnamefont
  {S.~D.}\ \bibnamefont {Huber}},\ }\bibfield  {title} {\bibinfo {title}
  {Observation of a phononic quadrupole topological insulator},\ }\href
  {https://doi.org/10.1038/nature25156} {\bibfield  {journal} {\bibinfo
  {journal} {Nature}\ }\textbf {\bibinfo {volume} {555}},\ \bibinfo {pages}
  {342} (\bibinfo {year} {2018})}\BibitemShut {NoStop}%
\bibitem [{\citenamefont {Bellec}\ \emph {et~al.}(2013)\citenamefont {Bellec},
  \citenamefont {Kuhl}, \citenamefont {Montambaux},\ and\ \citenamefont
  {Mortessagne}}]{Bellec2013}%
  \BibitemOpen
  \bibfield  {author} {\bibinfo {author} {\bibfnamefont {M.}~\bibnamefont
  {Bellec}}, \bibinfo {author} {\bibfnamefont {U.}~\bibnamefont {Kuhl}},
  \bibinfo {author} {\bibfnamefont {G.}~\bibnamefont {Montambaux}},\ and\
  \bibinfo {author} {\bibfnamefont {F.}~\bibnamefont {Mortessagne}},\
  }\bibfield  {title} {\bibinfo {title} {Topological transition of dirac points
  in a microwave experiment},\ }\href
  {https://doi.org/10.1103/PhysRevLett.110.033902} {\bibfield  {journal}
  {\bibinfo  {journal} {Phys. Rev. Lett.}\ }\textbf {\bibinfo {volume} {110}},\
  \bibinfo {pages} {033902} (\bibinfo {year} {2013})}\BibitemShut {NoStop}%
\bibitem [{\citenamefont {Hu}\ \emph {et~al.}(2015)\citenamefont {Hu},
  \citenamefont {Pillay}, \citenamefont {Wu}, \citenamefont {Pasek},
  \citenamefont {Shum},\ and\ \citenamefont {Chong}}]{Hu2015}%
  \BibitemOpen
  \bibfield  {author} {\bibinfo {author} {\bibfnamefont {W.}~\bibnamefont
  {Hu}}, \bibinfo {author} {\bibfnamefont {J.~C.}\ \bibnamefont {Pillay}},
  \bibinfo {author} {\bibfnamefont {K.}~\bibnamefont {Wu}}, \bibinfo {author}
  {\bibfnamefont {M.}~\bibnamefont {Pasek}}, \bibinfo {author} {\bibfnamefont
  {P.~P.}\ \bibnamefont {Shum}},\ and\ \bibinfo {author} {\bibfnamefont
  {Y.~D.}\ \bibnamefont {Chong}},\ }\bibfield  {title} {\bibinfo {title}
  {Measurement of a topological edge invariant in a microwave network},\ }\href
  {https://doi.org/10.1103/PhysRevX.5.011012} {\bibfield  {journal} {\bibinfo
  {journal} {Phys. Rev. X}\ }\textbf {\bibinfo {volume} {5}},\ \bibinfo {pages}
  {011012} (\bibinfo {year} {2015})}\BibitemShut {NoStop}%
\bibitem [{\citenamefont {Anderson}\ \emph {et~al.}(2016)\citenamefont
  {Anderson}, \citenamefont {Ma}, \citenamefont {Owens}, \citenamefont
  {Schuster},\ and\ \citenamefont {Simon}}]{Anderson2016}%
  \BibitemOpen
  \bibfield  {author} {\bibinfo {author} {\bibfnamefont {B.~M.}\ \bibnamefont
  {Anderson}}, \bibinfo {author} {\bibfnamefont {R.}~\bibnamefont {Ma}},
  \bibinfo {author} {\bibfnamefont {C.}~\bibnamefont {Owens}}, \bibinfo
  {author} {\bibfnamefont {D.~I.}\ \bibnamefont {Schuster}},\ and\ \bibinfo
  {author} {\bibfnamefont {J.}~\bibnamefont {Simon}},\ }\bibfield  {title}
  {\bibinfo {title} {Engineering topological many-body materials in microwave
  cavity arrays},\ }\href {https://doi.org/10.1103/PhysRevX.6.041043}
  {\bibfield  {journal} {\bibinfo  {journal} {Phys. Rev. X}\ }\textbf {\bibinfo
  {volume} {6}},\ \bibinfo {pages} {041043} (\bibinfo {year}
  {2016})}\BibitemShut {NoStop}%
\bibitem [{\citenamefont {Peterson}\ \emph {et~al.}(2018)\citenamefont
  {Peterson}, \citenamefont {Benalcazar}, \citenamefont {Hughes},\ and\
  \citenamefont {Bahl}}]{Peterson2018}%
  \BibitemOpen
  \bibfield  {author} {\bibinfo {author} {\bibfnamefont {C.~W.}\ \bibnamefont
  {Peterson}}, \bibinfo {author} {\bibfnamefont {W.~A.}\ \bibnamefont
  {Benalcazar}}, \bibinfo {author} {\bibfnamefont {T.~L.}\ \bibnamefont
  {Hughes}},\ and\ \bibinfo {author} {\bibfnamefont {G.}~\bibnamefont {Bahl}},\
  }\bibfield  {title} {\bibinfo {title} {A quantized microwave quadrupole
  insulator with topologically protected corner states},\ }\href
  {https://doi.org/10.1038/nature25777} {\bibfield  {journal} {\bibinfo
  {journal} {Nature}\ }\textbf {\bibinfo {volume} {555}},\ \bibinfo {pages}
  {346} (\bibinfo {year} {2018})}\BibitemShut {NoStop}%
\bibitem [{\citenamefont {Yu}\ \emph {et~al.}(2020)\citenamefont {Yu},
  \citenamefont {Song}, \citenamefont {Chen}, \citenamefont {Chen},
  \citenamefont {Ye}, \citenamefont {Shen}, \citenamefont {Cheng},\ and\
  \citenamefont {Li}}]{Yu2020}%
  \BibitemOpen
  \bibfield  {author} {\bibinfo {author} {\bibfnamefont {Y.}~\bibnamefont
  {Yu}}, \bibinfo {author} {\bibfnamefont {W.}~\bibnamefont {Song}}, \bibinfo
  {author} {\bibfnamefont {C.}~\bibnamefont {Chen}}, \bibinfo {author}
  {\bibfnamefont {T.}~\bibnamefont {Chen}}, \bibinfo {author} {\bibfnamefont
  {H.}~\bibnamefont {Ye}}, \bibinfo {author} {\bibfnamefont {X.}~\bibnamefont
  {Shen}}, \bibinfo {author} {\bibfnamefont {Q.}~\bibnamefont {Cheng}},\ and\
  \bibinfo {author} {\bibfnamefont {T.}~\bibnamefont {Li}},\ }\bibfield
  {title} {\bibinfo {title} {{Phase transition of non-Hermitian topological
  edge states in microwave regime}},\ }\href
  {https://doi.org/10.1063/5.0006144} {\bibfield  {journal} {\bibinfo
  {journal} {Appl. Phys. Lett.}\ }\textbf {\bibinfo {volume} {116}},\ \bibinfo
  {pages} {211104} (\bibinfo {year} {2020})}\BibitemShut {NoStop}%
\bibitem [{\citenamefont {Ma}\ and\ \citenamefont {Anlage}(2020)}]{Ma2020}%
  \BibitemOpen
  \bibfield  {author} {\bibinfo {author} {\bibfnamefont {S.}~\bibnamefont
  {Ma}}\ and\ \bibinfo {author} {\bibfnamefont {S.~M.}\ \bibnamefont
  {Anlage}},\ }\bibfield  {title} {\bibinfo {title} {{Microwave applications of
  photonic topological insulators}},\ }\href
  {https://doi.org/10.1063/5.0008046} {\bibfield  {journal} {\bibinfo
  {journal} {Appl. Phys. Lett.}\ }\textbf {\bibinfo {volume} {116}},\ \bibinfo
  {pages} {250502} (\bibinfo {year} {2020})}\BibitemShut {NoStop}%
\bibitem [{\citenamefont {Ningyuan}\ \emph {et~al.}(2015)\citenamefont
  {Ningyuan}, \citenamefont {Owens}, \citenamefont {Sommer}, \citenamefont
  {Schuster},\ and\ \citenamefont {Simon}}]{Jia2015}%
  \BibitemOpen
  \bibfield  {author} {\bibinfo {author} {\bibfnamefont {J.}~\bibnamefont
  {Ningyuan}}, \bibinfo {author} {\bibfnamefont {C.}~\bibnamefont {Owens}},
  \bibinfo {author} {\bibfnamefont {A.}~\bibnamefont {Sommer}}, \bibinfo
  {author} {\bibfnamefont {D.}~\bibnamefont {Schuster}},\ and\ \bibinfo
  {author} {\bibfnamefont {J.}~\bibnamefont {Simon}},\ }\bibfield  {title}
  {\bibinfo {title} {Time- and site-resolved dynamics in a topological
  circuit},\ }\href {https://doi.org/10.1103/PhysRevX.5.021031} {\bibfield
  {journal} {\bibinfo  {journal} {Phys. Rev. X}\ }\textbf {\bibinfo {volume}
  {5}},\ \bibinfo {pages} {021031} (\bibinfo {year} {2015})}\BibitemShut
  {NoStop}%
\bibitem [{\citenamefont {Albert}\ \emph {et~al.}(2015)\citenamefont {Albert},
  \citenamefont {Glazman},\ and\ \citenamefont {Jiang}}]{Albert2015}%
  \BibitemOpen
  \bibfield  {author} {\bibinfo {author} {\bibfnamefont {V.~V.}\ \bibnamefont
  {Albert}}, \bibinfo {author} {\bibfnamefont {L.~I.}\ \bibnamefont
  {Glazman}},\ and\ \bibinfo {author} {\bibfnamefont {L.}~\bibnamefont
  {Jiang}},\ }\bibfield  {title} {\bibinfo {title} {Topological properties of
  linear circuit lattices},\ }\href
  {https://doi.org/10.1103/PhysRevLett.114.173902} {\bibfield  {journal}
  {\bibinfo  {journal} {Phys. Rev. Lett.}\ }\textbf {\bibinfo {volume} {114}},\
  \bibinfo {pages} {173902} (\bibinfo {year} {2015})}\BibitemShut {NoStop}%
\bibitem [{\citenamefont {Lee}\ \emph {et~al.}(2018)\citenamefont {Lee},
  \citenamefont {Imhof}, \citenamefont {Berger}, \citenamefont {Bayer},
  \citenamefont {Brehm}, \citenamefont {Molenkamp}, \citenamefont {Kiessling},\
  and\ \citenamefont {Thomale}}]{Lee2018}%
  \BibitemOpen
  \bibfield  {author} {\bibinfo {author} {\bibfnamefont {C.~H.}\ \bibnamefont
  {Lee}}, \bibinfo {author} {\bibfnamefont {S.}~\bibnamefont {Imhof}}, \bibinfo
  {author} {\bibfnamefont {C.}~\bibnamefont {Berger}}, \bibinfo {author}
  {\bibfnamefont {F.}~\bibnamefont {Bayer}}, \bibinfo {author} {\bibfnamefont
  {J.}~\bibnamefont {Brehm}}, \bibinfo {author} {\bibfnamefont {L.~W.}\
  \bibnamefont {Molenkamp}}, \bibinfo {author} {\bibfnamefont {T.}~\bibnamefont
  {Kiessling}},\ and\ \bibinfo {author} {\bibfnamefont {R.}~\bibnamefont
  {Thomale}},\ }\bibfield  {title} {\bibinfo {title} {Topolectrical circuits},\
  }\href {https://doi.org/10.1038/s42005-018-0035-2} {\bibfield  {journal}
  {\bibinfo  {journal} {Commun. Phys.}\ }\textbf {\bibinfo {volume} {1}},\
  \bibinfo {pages} {39} (\bibinfo {year} {2018})}\BibitemShut {NoStop}%
\bibitem [{\citenamefont {Goren}\ \emph {et~al.}(2018)\citenamefont {Goren},
  \citenamefont {Plekhanov}, \citenamefont {Appas},\ and\ \citenamefont
  {Le~Hur}}]{Tal2018}%
  \BibitemOpen
  \bibfield  {author} {\bibinfo {author} {\bibfnamefont {T.}~\bibnamefont
  {Goren}}, \bibinfo {author} {\bibfnamefont {K.}~\bibnamefont {Plekhanov}},
  \bibinfo {author} {\bibfnamefont {F.}~\bibnamefont {Appas}},\ and\ \bibinfo
  {author} {\bibfnamefont {K.}~\bibnamefont {Le~Hur}},\ }\bibfield  {title}
  {\bibinfo {title} {Topological {Z}ak phase in strongly coupled {LC}
  circuits},\ }\href {https://doi.org/10.1103/PhysRevB.97.041106} {\bibfield
  {journal} {\bibinfo  {journal} {Phys. Rev. B}\ }\textbf {\bibinfo {volume}
  {97}},\ \bibinfo {pages} {041106} (\bibinfo {year} {2018})}\BibitemShut
  {NoStop}%
\bibitem [{\citenamefont {Li}\ \emph {et~al.}(2018)\citenamefont {Li},
  \citenamefont {Sun}, \citenamefont {Zhu}, \citenamefont {Guo}, \citenamefont
  {Jiang}, \citenamefont {Kariyado}, \citenamefont {Chen},\ and\ \citenamefont
  {Hu}}]{Li2018}%
  \BibitemOpen
  \bibfield  {author} {\bibinfo {author} {\bibfnamefont {Y.}~\bibnamefont
  {Li}}, \bibinfo {author} {\bibfnamefont {Y.}~\bibnamefont {Sun}}, \bibinfo
  {author} {\bibfnamefont {W.}~\bibnamefont {Zhu}}, \bibinfo {author}
  {\bibfnamefont {Z.}~\bibnamefont {Guo}}, \bibinfo {author} {\bibfnamefont
  {J.}~\bibnamefont {Jiang}}, \bibinfo {author} {\bibfnamefont
  {T.}~\bibnamefont {Kariyado}}, \bibinfo {author} {\bibfnamefont
  {H.}~\bibnamefont {Chen}},\ and\ \bibinfo {author} {\bibfnamefont
  {X.}~\bibnamefont {Hu}},\ }\bibfield  {title} {\bibinfo {title} {Topological
  {LC}-circuits based on microstrips and observation of electromagnetic modes
  with orbital angular momentum},\ }\href
  {https://doi.org/10.1038/s41467-018-07084-2} {\bibfield  {journal} {\bibinfo
  {journal} {Nat. Commun.}\ }\textbf {\bibinfo {volume} {9}},\ \bibinfo {pages}
  {4598} (\bibinfo {year} {2018})}\BibitemShut {NoStop}%
\bibitem [{\citenamefont {Hofmann}\ \emph {et~al.}(2019)\citenamefont
  {Hofmann}, \citenamefont {Helbig}, \citenamefont {Lee}, \citenamefont
  {Greiter},\ and\ \citenamefont {Thomale}}]{Hofmann2019}%
  \BibitemOpen
  \bibfield  {author} {\bibinfo {author} {\bibfnamefont {T.}~\bibnamefont
  {Hofmann}}, \bibinfo {author} {\bibfnamefont {T.}~\bibnamefont {Helbig}},
  \bibinfo {author} {\bibfnamefont {C.~H.}\ \bibnamefont {Lee}}, \bibinfo
  {author} {\bibfnamefont {M.}~\bibnamefont {Greiter}},\ and\ \bibinfo {author}
  {\bibfnamefont {R.}~\bibnamefont {Thomale}},\ }\bibfield  {title} {\bibinfo
  {title} {Chiral voltage propagation and calibration in a topolectrical
  {C}hern circuit},\ }\href {https://doi.org/10.1103/PhysRevLett.122.247702}
  {\bibfield  {journal} {\bibinfo  {journal} {Phys. Rev. Lett.}\ }\textbf
  {\bibinfo {volume} {122}},\ \bibinfo {pages} {247702} (\bibinfo {year}
  {2019})}\BibitemShut {NoStop}%
\bibitem [{\citenamefont {Haenel}\ \emph {et~al.}(2019)\citenamefont {Haenel},
  \citenamefont {Branch},\ and\ \citenamefont {Franz}}]{Haenel2019}%
  \BibitemOpen
  \bibfield  {author} {\bibinfo {author} {\bibfnamefont {R.}~\bibnamefont
  {Haenel}}, \bibinfo {author} {\bibfnamefont {T.}~\bibnamefont {Branch}},\
  and\ \bibinfo {author} {\bibfnamefont {M.}~\bibnamefont {Franz}},\ }\bibfield
   {title} {\bibinfo {title} {Chern insulators for electromagnetic waves in
  electrical circuit networks},\ }\href
  {https://doi.org/10.1103/PhysRevB.99.235110} {\bibfield  {journal} {\bibinfo
  {journal} {Phys. Rev. B}\ }\textbf {\bibinfo {volume} {99}},\ \bibinfo
  {pages} {235110} (\bibinfo {year} {2019})}\BibitemShut {NoStop}%
\bibitem [{\citenamefont {Zhu}\ \emph {et~al.}(2019)\citenamefont {Zhu},
  \citenamefont {Long}, \citenamefont {Chen},\ and\ \citenamefont
  {Ren}}]{Zhu2019}%
  \BibitemOpen
  \bibfield  {author} {\bibinfo {author} {\bibfnamefont {W.}~\bibnamefont
  {Zhu}}, \bibinfo {author} {\bibfnamefont {Y.}~\bibnamefont {Long}}, \bibinfo
  {author} {\bibfnamefont {H.}~\bibnamefont {Chen}},\ and\ \bibinfo {author}
  {\bibfnamefont {J.}~\bibnamefont {Ren}},\ }\bibfield  {title} {\bibinfo
  {title} {Quantum valley {H}all effects and spin-valley locking in topological
  {K}ane-{M}ele circuit networks},\ }\href
  {https://doi.org/10.1103/PhysRevB.99.115410} {\bibfield  {journal} {\bibinfo
  {journal} {Phys. Rev. B}\ }\textbf {\bibinfo {volume} {99}},\ \bibinfo
  {pages} {115410} (\bibinfo {year} {2019})}\BibitemShut {NoStop}%
\bibitem [{\citenamefont {Helbig}\ \emph {et~al.}(2020)\citenamefont {Helbig},
  \citenamefont {Hofmann}, \citenamefont {Imhof}, \citenamefont {Abdelghany},
  \citenamefont {Kiessling}, \citenamefont {Molenkamp}, \citenamefont {Lee},
  \citenamefont {Szameit}, \citenamefont {Greiter},\ and\ \citenamefont
  {Thomale}}]{Helbig2020}%
  \BibitemOpen
  \bibfield  {author} {\bibinfo {author} {\bibfnamefont {T.}~\bibnamefont
  {Helbig}}, \bibinfo {author} {\bibfnamefont {T.}~\bibnamefont {Hofmann}},
  \bibinfo {author} {\bibfnamefont {S.}~\bibnamefont {Imhof}}, \bibinfo
  {author} {\bibfnamefont {M.}~\bibnamefont {Abdelghany}}, \bibinfo {author}
  {\bibfnamefont {T.}~\bibnamefont {Kiessling}}, \bibinfo {author}
  {\bibfnamefont {L.~W.}\ \bibnamefont {Molenkamp}}, \bibinfo {author}
  {\bibfnamefont {C.~H.}\ \bibnamefont {Lee}}, \bibinfo {author} {\bibfnamefont
  {A.}~\bibnamefont {Szameit}}, \bibinfo {author} {\bibfnamefont
  {M.}~\bibnamefont {Greiter}},\ and\ \bibinfo {author} {\bibfnamefont
  {R.}~\bibnamefont {Thomale}},\ }\bibfield  {title} {\bibinfo {title}
  {Generalized bulk-boundary correspondence in non-{H}ermitian topolectrical
  circuits},\ }\href {https://doi.org/10.1038/s41567-020-0922-9} {\bibfield
  {journal} {\bibinfo  {journal} {Nature Phys.}\ }\textbf {\bibinfo {volume}
  {16}},\ \bibinfo {pages} {747} (\bibinfo {year} {2020})}\BibitemShut
  {NoStop}%
\bibitem [{\citenamefont {Hofmann}\ \emph {et~al.}(2020)\citenamefont
  {Hofmann}, \citenamefont {Helbig}, \citenamefont {Schindler}, \citenamefont
  {Salgo}, \citenamefont {Brzezi\ifmmode~\acute{n}\else \'{n}\fi{}ska},
  \citenamefont {Greiter}, \citenamefont {Kiessling}, \citenamefont {Wolf},
  \citenamefont {Vollhardt}, \citenamefont {Kaba\ifmmode~\check{s}\else
  \v{s}\fi{}i}, \citenamefont {Lee}, \citenamefont {Bilu\ifmmode \check{s}\else
  \v{s}\fi{}i\ifmmode~\acute{c}\else \'{c}\fi{}}, \citenamefont {Thomale},\
  and\ \citenamefont {Neupert}}]{Hofmann2020}%
  \BibitemOpen
  \bibfield  {author} {\bibinfo {author} {\bibfnamefont {T.}~\bibnamefont
  {Hofmann}}, \bibinfo {author} {\bibfnamefont {T.}~\bibnamefont {Helbig}},
  \bibinfo {author} {\bibfnamefont {F.}~\bibnamefont {Schindler}}, \bibinfo
  {author} {\bibfnamefont {N.}~\bibnamefont {Salgo}}, \bibinfo {author}
  {\bibfnamefont {M.}~\bibnamefont {Brzezi\ifmmode~\acute{n}\else
  \'{n}\fi{}ska}}, \bibinfo {author} {\bibfnamefont {M.}~\bibnamefont
  {Greiter}}, \bibinfo {author} {\bibfnamefont {T.}~\bibnamefont {Kiessling}},
  \bibinfo {author} {\bibfnamefont {D.}~\bibnamefont {Wolf}}, \bibinfo {author}
  {\bibfnamefont {A.}~\bibnamefont {Vollhardt}}, \bibinfo {author}
  {\bibfnamefont {A.}~\bibnamefont {Kaba\ifmmode~\check{s}\else \v{s}\fi{}i}},
  \bibinfo {author} {\bibfnamefont {C.~H.}\ \bibnamefont {Lee}}, \bibinfo
  {author} {\bibfnamefont {A.}~\bibnamefont {Bilu\ifmmode \check{s}\else
  \v{s}\fi{}i\ifmmode~\acute{c}\else \'{c}\fi{}}}, \bibinfo {author}
  {\bibfnamefont {R.}~\bibnamefont {Thomale}},\ and\ \bibinfo {author}
  {\bibfnamefont {T.}~\bibnamefont {Neupert}},\ }\bibfield  {title} {\bibinfo
  {title} {Reciprocal skin effect and its realization in a topolectrical
  circuit},\ }\href {https://doi.org/10.1103/PhysRevResearch.2.023265}
  {\bibfield  {journal} {\bibinfo  {journal} {Phys. Rev. Res.}\ }\textbf
  {\bibinfo {volume} {2}},\ \bibinfo {pages} {023265} (\bibinfo {year}
  {2020})}\BibitemShut {NoStop}%
\bibitem [{\citenamefont {Wang}\ \emph {et~al.}(2020)\citenamefont {Wang},
  \citenamefont {Price}, \citenamefont {Zhang},\ and\ \citenamefont
  {Chong}}]{Wang2020}%
  \BibitemOpen
  \bibfield  {author} {\bibinfo {author} {\bibfnamefont {Y.}~\bibnamefont
  {Wang}}, \bibinfo {author} {\bibfnamefont {H.~M.}\ \bibnamefont {Price}},
  \bibinfo {author} {\bibfnamefont {B.}~\bibnamefont {Zhang}},\ and\ \bibinfo
  {author} {\bibfnamefont {Y.~D.}\ \bibnamefont {Chong}},\ }\bibfield  {title}
  {\bibinfo {title} {Circuit implementation of a four-dimensional topological
  insulator},\ }\href {https://doi.org/10.1038/s41467-020-15940-3} {\bibfield
  {journal} {\bibinfo  {journal} {Nat. Commun.}\ }\textbf {\bibinfo {volume}
  {11}},\ \bibinfo {pages} {2356} (\bibinfo {year} {2020})}\BibitemShut
  {NoStop}%
\bibitem [{\citenamefont {Rafi-Ul-Islam}\ \emph {et~al.}(2020)\citenamefont
  {Rafi-Ul-Islam}, \citenamefont {Bin~Siu},\ and\ \citenamefont
  {Jalil}}]{Rafi-Ul-Islam2020}%
  \BibitemOpen
  \bibfield  {author} {\bibinfo {author} {\bibfnamefont {S.~M.}\ \bibnamefont
  {Rafi-Ul-Islam}}, \bibinfo {author} {\bibfnamefont {Z.}~\bibnamefont
  {Bin~Siu}},\ and\ \bibinfo {author} {\bibfnamefont {M.~B.~A.}\ \bibnamefont
  {Jalil}},\ }\bibfield  {title} {\bibinfo {title} {Topoelectrical circuit
  realization of a {W}eyl semimetal heterojunction},\ }\href
  {https://doi.org/10.1038/s42005-020-0336-0} {\bibfield  {journal} {\bibinfo
  {journal} {Commun. Phys.}\ }\textbf {\bibinfo {volume} {3}},\ \bibinfo
  {pages} {72} (\bibinfo {year} {2020})}\BibitemShut {NoStop}%
\bibitem [{\citenamefont {Dong}\ \emph {et~al.}(2021)\citenamefont {Dong},
  \citenamefont {Juri\ifmmode \check{c}\else \v{c}\fi{}i\ifmmode~\acute{c}\else
  \'{c}\fi{}},\ and\ \citenamefont {Roy}}]{Dong2021}%
  \BibitemOpen
  \bibfield  {author} {\bibinfo {author} {\bibfnamefont {J.}~\bibnamefont
  {Dong}}, \bibinfo {author} {\bibfnamefont {V.}~\bibnamefont {Juri\ifmmode
  \check{c}\else \v{c}\fi{}i\ifmmode~\acute{c}\else \'{c}\fi{}}},\ and\
  \bibinfo {author} {\bibfnamefont {B.}~\bibnamefont {Roy}},\ }\bibfield
  {title} {\bibinfo {title} {Topolectric circuits: Theory and construction},\
  }\href {https://doi.org/10.1103/PhysRevResearch.3.023056} {\bibfield
  {journal} {\bibinfo  {journal} {Phys. Rev. Res.}\ }\textbf {\bibinfo {volume}
  {3}},\ \bibinfo {pages} {023056} (\bibinfo {year} {2021})}\BibitemShut
  {NoStop}%
\bibitem [{\citenamefont {Kotwal}\ \emph {et~al.}(2021)\citenamefont {Kotwal},
  \citenamefont {Moseley}, \citenamefont {Stegmaier}, \citenamefont {Imhof},
  \citenamefont {Brand}, \citenamefont {Kießling}, \citenamefont {Thomale},
  \citenamefont {Ronellenfitsch},\ and\ \citenamefont {Dunkel}}]{Kotwal2021}%
  \BibitemOpen
  \bibfield  {author} {\bibinfo {author} {\bibfnamefont {T.}~\bibnamefont
  {Kotwal}}, \bibinfo {author} {\bibfnamefont {F.}~\bibnamefont {Moseley}},
  \bibinfo {author} {\bibfnamefont {A.}~\bibnamefont {Stegmaier}}, \bibinfo
  {author} {\bibfnamefont {S.}~\bibnamefont {Imhof}}, \bibinfo {author}
  {\bibfnamefont {H.}~\bibnamefont {Brand}}, \bibinfo {author} {\bibfnamefont
  {T.}~\bibnamefont {Kießling}}, \bibinfo {author} {\bibfnamefont
  {R.}~\bibnamefont {Thomale}}, \bibinfo {author} {\bibfnamefont
  {H.}~\bibnamefont {Ronellenfitsch}},\ and\ \bibinfo {author} {\bibfnamefont
  {J.}~\bibnamefont {Dunkel}},\ }\bibfield  {title} {\bibinfo {title} {Active
  topolectrical circuits},\ }\href {https://doi.org/10.1073/pnas.2106411118}
  {\bibfield  {journal} {\bibinfo  {journal} {PNAS}\ }\textbf {\bibinfo
  {volume} {118}},\ \bibinfo {pages} {e2106411118} (\bibinfo {year}
  {2021})}\BibitemShut {NoStop}%
\bibitem [{\citenamefont {Yang}\ \emph {et~al.}(2023)\citenamefont {Yang},
  \citenamefont {Song}, \citenamefont {Cao},\ and\ \citenamefont
  {Yan}}]{Yang2023}%
  \BibitemOpen
  \bibfield  {author} {\bibinfo {author} {\bibfnamefont {H.}~\bibnamefont
  {Yang}}, \bibinfo {author} {\bibfnamefont {L.}~\bibnamefont {Song}}, \bibinfo
  {author} {\bibfnamefont {Y.}~\bibnamefont {Cao}},\ and\ \bibinfo {author}
  {\bibfnamefont {P.}~\bibnamefont {Yan}},\ }\bibfield  {title} {\bibinfo
  {title} {Realization of {W}ilson fermions in topolectrical circuits},\ }\href
  {https://doi.org/10.1038/s42005-023-01326-6} {\bibfield  {journal} {\bibinfo
  {journal} {Commun. Phys.}\ }\textbf {\bibinfo {volume} {6}},\ \bibinfo
  {pages} {211} (\bibinfo {year} {2023})}\BibitemShut {NoStop}%
\bibitem [{\citenamefont {Zheng}\ \emph {et~al.}(2022)\citenamefont {Zheng},
  \citenamefont {Chen},\ and\ \citenamefont {Zhang}}]{Zheng2022}%
  \BibitemOpen
  \bibfield  {author} {\bibinfo {author} {\bibfnamefont {X.}~\bibnamefont
  {Zheng}}, \bibinfo {author} {\bibfnamefont {T.}~\bibnamefont {Chen}},\ and\
  \bibinfo {author} {\bibfnamefont {X.}~\bibnamefont {Zhang}},\ }\bibfield
  {title} {\bibinfo {title} {Topolectrical circuit realization of quadrupolar
  surface semimetals},\ }\href {https://doi.org/10.1103/PhysRevB.106.035308}
  {\bibfield  {journal} {\bibinfo  {journal} {Phys. Rev. B}\ }\textbf {\bibinfo
  {volume} {106}},\ \bibinfo {pages} {035308} (\bibinfo {year}
  {2022})}\BibitemShut {NoStop}%
\bibitem [{\citenamefont {Zhang}\ \emph {et~al.}(2023)\citenamefont {Zhang},
  \citenamefont {Chen}, \citenamefont {Li}, \citenamefont {Lee},\ and\
  \citenamefont {Zhang}}]{Zhang2023}%
  \BibitemOpen
  \bibfield  {author} {\bibinfo {author} {\bibfnamefont {H.}~\bibnamefont
  {Zhang}}, \bibinfo {author} {\bibfnamefont {T.}~\bibnamefont {Chen}},
  \bibinfo {author} {\bibfnamefont {L.}~\bibnamefont {Li}}, \bibinfo {author}
  {\bibfnamefont {C.~H.}\ \bibnamefont {Lee}},\ and\ \bibinfo {author}
  {\bibfnamefont {X.}~\bibnamefont {Zhang}},\ }\bibfield  {title} {\bibinfo
  {title} {Electrical circuit realization of topological switching for the
  non-{H}ermitian skin effect},\ }\href
  {https://doi.org/10.1103/PhysRevB.107.085426} {\bibfield  {journal} {\bibinfo
   {journal} {Phys. Rev. B}\ }\textbf {\bibinfo {volume} {107}},\ \bibinfo
  {pages} {085426} (\bibinfo {year} {2023})}\BibitemShut {NoStop}%
\bibitem [{\citenamefont {Imhof}\ \emph {et~al.}(2018)\citenamefont {Imhof},
  \citenamefont {Berger}, \citenamefont {Bayer}, \citenamefont {Brehm},
  \citenamefont {Molenkamp}, \citenamefont {Kiessling}, \citenamefont
  {Schindler}, \citenamefont {Lee}, \citenamefont {Greiter}, \citenamefont
  {Neupert},\ and\ \citenamefont {Thomale}}]{Imhof2018}%
  \BibitemOpen
  \bibfield  {author} {\bibinfo {author} {\bibfnamefont {S.}~\bibnamefont
  {Imhof}}, \bibinfo {author} {\bibfnamefont {C.}~\bibnamefont {Berger}},
  \bibinfo {author} {\bibfnamefont {F.}~\bibnamefont {Bayer}}, \bibinfo
  {author} {\bibfnamefont {J.}~\bibnamefont {Brehm}}, \bibinfo {author}
  {\bibfnamefont {L.~W.}\ \bibnamefont {Molenkamp}}, \bibinfo {author}
  {\bibfnamefont {T.}~\bibnamefont {Kiessling}}, \bibinfo {author}
  {\bibfnamefont {F.}~\bibnamefont {Schindler}}, \bibinfo {author}
  {\bibfnamefont {C.~H.}\ \bibnamefont {Lee}}, \bibinfo {author} {\bibfnamefont
  {M.}~\bibnamefont {Greiter}}, \bibinfo {author} {\bibfnamefont
  {T.}~\bibnamefont {Neupert}},\ and\ \bibinfo {author} {\bibfnamefont
  {R.}~\bibnamefont {Thomale}},\ }\bibfield  {title} {\bibinfo {title}
  {Topolectrical-circuit realization of topological corner modes},\ }\href
  {https://doi.org/10.1038/s41567-018-0246-1} {\bibfield  {journal} {\bibinfo
  {journal} {Nature Phys.}\ }\textbf {\bibinfo {volume} {14}},\ \bibinfo
  {pages} {925} (\bibinfo {year} {2018})}\BibitemShut {NoStop}%
\bibitem [{\citenamefont {Stegmaier}\ \emph {et~al.}(2021)\citenamefont
  {Stegmaier}, \citenamefont {Imhof}, \citenamefont {Helbig}, \citenamefont
  {Hofmann}, \citenamefont {Lee}, \citenamefont {Kremer}, \citenamefont
  {Fritzsche}, \citenamefont {Feichtner}, \citenamefont {Klembt}, \citenamefont
  {H\"ofling}, \citenamefont {Boettcher}, \citenamefont {Fulga}, \citenamefont
  {Ma}, \citenamefont {Schmidt}, \citenamefont {Greiter}, \citenamefont
  {Kiessling}, \citenamefont {Szameit},\ and\ \citenamefont
  {Thomale}}]{Stegmaier2021}%
  \BibitemOpen
  \bibfield  {author} {\bibinfo {author} {\bibfnamefont {A.}~\bibnamefont
  {Stegmaier}}, \bibinfo {author} {\bibfnamefont {S.}~\bibnamefont {Imhof}},
  \bibinfo {author} {\bibfnamefont {T.}~\bibnamefont {Helbig}}, \bibinfo
  {author} {\bibfnamefont {T.}~\bibnamefont {Hofmann}}, \bibinfo {author}
  {\bibfnamefont {C.~H.}\ \bibnamefont {Lee}}, \bibinfo {author} {\bibfnamefont
  {M.}~\bibnamefont {Kremer}}, \bibinfo {author} {\bibfnamefont
  {A.}~\bibnamefont {Fritzsche}}, \bibinfo {author} {\bibfnamefont
  {T.}~\bibnamefont {Feichtner}}, \bibinfo {author} {\bibfnamefont
  {S.}~\bibnamefont {Klembt}}, \bibinfo {author} {\bibfnamefont
  {S.}~\bibnamefont {H\"ofling}}, \bibinfo {author} {\bibfnamefont
  {I.}~\bibnamefont {Boettcher}}, \bibinfo {author} {\bibfnamefont {I.~C.}\
  \bibnamefont {Fulga}}, \bibinfo {author} {\bibfnamefont {L.}~\bibnamefont
  {Ma}}, \bibinfo {author} {\bibfnamefont {O.~G.}\ \bibnamefont {Schmidt}},
  \bibinfo {author} {\bibfnamefont {M.}~\bibnamefont {Greiter}}, \bibinfo
  {author} {\bibfnamefont {T.}~\bibnamefont {Kiessling}}, \bibinfo {author}
  {\bibfnamefont {A.}~\bibnamefont {Szameit}},\ and\ \bibinfo {author}
  {\bibfnamefont {R.}~\bibnamefont {Thomale}},\ }\bibfield  {title} {\bibinfo
  {title} {Topological defect engineering and $\mathcal{P}\mathcal{T}$ symmetry
  in non-{H}ermitian electrical circuits},\ }\href
  {https://doi.org/10.1103/PhysRevLett.126.215302} {\bibfield  {journal}
  {\bibinfo  {journal} {Phys. Rev. Lett.}\ }\textbf {\bibinfo {volume} {126}},\
  \bibinfo {pages} {215302} (\bibinfo {year} {2021})}\BibitemShut {NoStop}%
\bibitem [{\citenamefont {Lenz}\ and\ \citenamefont
  {Stollmann}(2003)}]{Lenz2003}%
  \BibitemOpen
  \bibfield  {author} {\bibinfo {author} {\bibfnamefont {D.}~\bibnamefont
  {Lenz}}\ and\ \bibinfo {author} {\bibfnamefont {P.}~\bibnamefont
  {Stollmann}},\ }\bibfield  {title} {\bibinfo {title} {Aperiodic order and
  quasicrystals: Spectral properties},\ }\href
  {https://doi.org/10.1007/s00023-003-0973-3} {\bibfield  {journal} {\bibinfo
  {journal} {Ann. Henri Poincaré}\ }\textbf {\bibinfo {volume} {4}},\ \bibinfo
  {pages} {933} (\bibinfo {year} {2003})}\BibitemShut {NoStop}%
\bibitem [{\citenamefont {Pikovsky}\ \emph {et~al.}(1995)\citenamefont
  {Pikovsky}, \citenamefont {Zaks}, \citenamefont {Feudel},\ and\ \citenamefont
  {Kurths}}]{Pikovsky1995}%
  \BibitemOpen
  \bibfield  {author} {\bibinfo {author} {\bibfnamefont {A.~S.}\ \bibnamefont
  {Pikovsky}}, \bibinfo {author} {\bibfnamefont {M.~A.}\ \bibnamefont {Zaks}},
  \bibinfo {author} {\bibfnamefont {U.}~\bibnamefont {Feudel}},\ and\ \bibinfo
  {author} {\bibfnamefont {J.}~\bibnamefont {Kurths}},\ }\bibfield  {title}
  {\bibinfo {title} {Singular continuous spectra in dissipative dynamics},\
  }\href {https://doi.org/10.1103/PhysRevE.52.285} {\bibfield  {journal}
  {\bibinfo  {journal} {Phys. Rev. E}\ }\textbf {\bibinfo {volume} {52}},\
  \bibinfo {pages} {285} (\bibinfo {year} {1995})}\BibitemShut {NoStop}%
\bibitem [{\citenamefont {Jagannathan}(2021)}]{Jagannathan2021}%
  \BibitemOpen
  \bibfield  {author} {\bibinfo {author} {\bibfnamefont {A.}~\bibnamefont
  {Jagannathan}},\ }\bibfield  {title} {\bibinfo {title} {The {F}ibonacci
  quasicrystal: Case study of hidden dimensions and multifractality},\ }\href
  {https://doi.org/10.1103/RevModPhys.93.045001} {\bibfield  {journal}
  {\bibinfo  {journal} {Rev. Mod. Phys.}\ }\textbf {\bibinfo {volume} {93}},\
  \bibinfo {pages} {045001} (\bibinfo {year} {2021})}\BibitemShut {NoStop}%
\bibitem [{\citenamefont {Tanese}\ \emph {et~al.}(2014)\citenamefont {Tanese},
  \citenamefont {Gurevich}, \citenamefont {Baboux}, \citenamefont {Jacqmin},
  \citenamefont {Lema\^{\i}tre}, \citenamefont {Galopin}, \citenamefont
  {Sagnes}, \citenamefont {Amo}, \citenamefont {Bloch},\ and\ \citenamefont
  {Akkermans}}]{Tanese2014}%
  \BibitemOpen
  \bibfield  {author} {\bibinfo {author} {\bibfnamefont {D.}~\bibnamefont
  {Tanese}}, \bibinfo {author} {\bibfnamefont {E.}~\bibnamefont {Gurevich}},
  \bibinfo {author} {\bibfnamefont {F.}~\bibnamefont {Baboux}}, \bibinfo
  {author} {\bibfnamefont {T.}~\bibnamefont {Jacqmin}}, \bibinfo {author}
  {\bibfnamefont {A.}~\bibnamefont {Lema\^{\i}tre}}, \bibinfo {author}
  {\bibfnamefont {E.}~\bibnamefont {Galopin}}, \bibinfo {author} {\bibfnamefont
  {I.}~\bibnamefont {Sagnes}}, \bibinfo {author} {\bibfnamefont
  {A.}~\bibnamefont {Amo}}, \bibinfo {author} {\bibfnamefont {J.}~\bibnamefont
  {Bloch}},\ and\ \bibinfo {author} {\bibfnamefont {E.}~\bibnamefont
  {Akkermans}},\ }\bibfield  {title} {\bibinfo {title} {Fractal energy spectrum
  of a polariton gas in a {F}ibonacci quasiperiodic potential},\ }\href
  {https://doi.org/10.1103/PhysRevLett.112.146404} {\bibfield  {journal}
  {\bibinfo  {journal} {Phys. Rev. Lett.}\ }\textbf {\bibinfo {volume} {112}},\
  \bibinfo {pages} {146404} (\bibinfo {year} {2014})}\BibitemShut {NoStop}%
\bibitem [{\citenamefont {Kraus}\ \emph {et~al.}(2012)\citenamefont {Kraus},
  \citenamefont {Lahini}, \citenamefont {Ringel}, \citenamefont {Verbin},\ and\
  \citenamefont {Zilberberg}}]{Kraus2012}%
  \BibitemOpen
  \bibfield  {author} {\bibinfo {author} {\bibfnamefont {Y.~E.}\ \bibnamefont
  {Kraus}}, \bibinfo {author} {\bibfnamefont {Y.}~\bibnamefont {Lahini}},
  \bibinfo {author} {\bibfnamefont {Z.}~\bibnamefont {Ringel}}, \bibinfo
  {author} {\bibfnamefont {M.}~\bibnamefont {Verbin}},\ and\ \bibinfo {author}
  {\bibfnamefont {O.}~\bibnamefont {Zilberberg}},\ }\bibfield  {title}
  {\bibinfo {title} {Topological states and adiabatic pumping in
  quasicrystals},\ }\href {https://doi.org/10.1103/PhysRevLett.109.106402}
  {\bibfield  {journal} {\bibinfo  {journal} {Phys. Rev. Lett.}\ }\textbf
  {\bibinfo {volume} {109}},\ \bibinfo {pages} {106402} (\bibinfo {year}
  {2012})}\BibitemShut {NoStop}%
\bibitem [{\citenamefont {Verbin}\ \emph {et~al.}(2013)\citenamefont {Verbin},
  \citenamefont {Zilberberg}, \citenamefont {Kraus}, \citenamefont {Lahini},\
  and\ \citenamefont {Silberberg}}]{Verbin2013}%
  \BibitemOpen
  \bibfield  {author} {\bibinfo {author} {\bibfnamefont {M.}~\bibnamefont
  {Verbin}}, \bibinfo {author} {\bibfnamefont {O.}~\bibnamefont {Zilberberg}},
  \bibinfo {author} {\bibfnamefont {Y.~E.}\ \bibnamefont {Kraus}}, \bibinfo
  {author} {\bibfnamefont {Y.}~\bibnamefont {Lahini}},\ and\ \bibinfo {author}
  {\bibfnamefont {Y.}~\bibnamefont {Silberberg}},\ }\bibfield  {title}
  {\bibinfo {title} {Observation of topological phase transitions in photonic
  quasicrystals},\ }\href {https://doi.org/10.1103/PhysRevLett.110.076403}
  {\bibfield  {journal} {\bibinfo  {journal} {Phys. Rev. Lett.}\ }\textbf
  {\bibinfo {volume} {110}},\ \bibinfo {pages} {076403} (\bibinfo {year}
  {2013})}\BibitemShut {NoStop}%
\bibitem [{\citenamefont {Stegmaier}\ \emph {et~al.}(2023)\citenamefont
  {Stegmaier}, \citenamefont {Brand}, \citenamefont {Imhof}, \citenamefont
  {Fritzsche}, \citenamefont {Helbig}, \citenamefont {Hofmann}, \citenamefont
  {Boettcher}, \citenamefont {Greiter}, \citenamefont {Lee}, \citenamefont
  {Bahl}, \citenamefont {Szameit}, \citenamefont {Kießling}, \citenamefont
  {Thomale},\ and\ \citenamefont {Upreti}}]{Stegmaier2023}%
  \BibitemOpen
  \bibfield  {author} {\bibinfo {author} {\bibfnamefont {A.}~\bibnamefont
  {Stegmaier}}, \bibinfo {author} {\bibfnamefont {H.}~\bibnamefont {Brand}},
  \bibinfo {author} {\bibfnamefont {S.}~\bibnamefont {Imhof}}, \bibinfo
  {author} {\bibfnamefont {A.}~\bibnamefont {Fritzsche}}, \bibinfo {author}
  {\bibfnamefont {T.}~\bibnamefont {Helbig}}, \bibinfo {author} {\bibfnamefont
  {T.}~\bibnamefont {Hofmann}}, \bibinfo {author} {\bibfnamefont
  {I.}~\bibnamefont {Boettcher}}, \bibinfo {author} {\bibfnamefont
  {M.}~\bibnamefont {Greiter}}, \bibinfo {author} {\bibfnamefont {C.~H.}\
  \bibnamefont {Lee}}, \bibinfo {author} {\bibfnamefont {G.}~\bibnamefont
  {Bahl}}, \bibinfo {author} {\bibfnamefont {A.}~\bibnamefont {Szameit}},
  \bibinfo {author} {\bibfnamefont {T.}~\bibnamefont {Kießling}}, \bibinfo
  {author} {\bibfnamefont {R.}~\bibnamefont {Thomale}},\ and\ \bibinfo {author}
  {\bibfnamefont {L.~K.}\ \bibnamefont {Upreti}},\ }\bibfield  {title}
  {\bibinfo {title} {Realizing efficient topological temporal pumping in
  electrical circuits},\ }\href {https://arxiv.org/abs/2306.15434} {\bibfield
  {journal} {\bibinfo  {journal} {Preprint at
  https://arxiv.org/abs/2306.15434}\ } (\bibinfo {year} {2023})}\BibitemShut
  {NoStop}%
\bibitem [{\citenamefont {Zou}\ \emph {et~al.}(2023)\citenamefont {Zou},
  \citenamefont {Chen}, \citenamefont {Meng}, \citenamefont {Ang},
  \citenamefont {Zhang},\ and\ \citenamefont {Lee}}]{Zou2023}%
  \BibitemOpen
  \bibfield  {author} {\bibinfo {author} {\bibfnamefont {D.}~\bibnamefont
  {Zou}}, \bibinfo {author} {\bibfnamefont {T.}~\bibnamefont {Chen}}, \bibinfo
  {author} {\bibfnamefont {H.}~\bibnamefont {Meng}}, \bibinfo {author}
  {\bibfnamefont {Y.~S.}\ \bibnamefont {Ang}}, \bibinfo {author} {\bibfnamefont
  {X.}~\bibnamefont {Zhang}},\ and\ \bibinfo {author} {\bibfnamefont {C.~H.}\
  \bibnamefont {Lee}},\ }\bibfield  {title} {\bibinfo {title} {Experimental
  observation of exceptional bound states in a classical circuit network},\
  }\href {https://arxiv.org/abs/2308.01970} {\bibfield  {journal} {\bibinfo
  {journal} {Preprint at https://arxiv.org/abs/2308.01970}\ } (\bibinfo {year}
  {2023})}\BibitemShut {NoStop}%
\bibitem [{\citenamefont {Sütő}(1989)}]{Suetoe1989}%
  \BibitemOpen
  \bibfield  {author} {\bibinfo {author} {\bibfnamefont {A.}~\bibnamefont
  {Sütő}},\ }\bibfield  {title} {\bibinfo {title} {Singular continuous
  spectrum on a {C}antor set of zero {L}ebesgue measure for the {F}ibonacci
  {H}amiltonian},\ }\href {https://doi.org/10.1007/BF01044450} {\bibfield
  {journal} {\bibinfo  {journal} {J. Stat. Phys.}\ }\textbf {\bibinfo {volume}
  {56}},\ \bibinfo {pages} {525} (\bibinfo {year} {1989})}\BibitemShut
  {NoStop}%
\bibitem [{SM()}]{SM}%
  \BibitemOpen
  \href@noop {} {\bibinfo  {journal} {In the Supplemental Material, we show
  details of the experimental realization before discussing how realistic
  capacitors and inductors affect the two point impedance. We proceed with
  discussing the matrix pencil method, before showing reconstructed spectra for
  the system under periodic boundary conditions.}\ }\BibitemShut {NoStop}%
\bibitem [{\citenamefont {O'Haver}(2023)}]{OHaverBook}%
  \BibitemOpen
\bibfield  {journal} {  }\bibfield  {author} {\bibinfo {author} {\bibfnamefont
  {T.}~\bibnamefont {O'Haver}},\ }\href@noop {} {\emph {\bibinfo {title} {A
  Pragmatic Introduction to Signal Processing 2023: with applications in
  scientific measurement}}},\ McGraw-Hill series in electrical engineering\
  (\bibinfo  {publisher} {Independently Published},\ \bibinfo {year}
  {2023})\BibitemShut {NoStop}%
\bibitem [{\citenamefont {Butterworth}(1930)}]{Butterworth1930}%
  \BibitemOpen
  \bibfield  {author} {\bibinfo {author} {\bibfnamefont {S.}~\bibnamefont
  {Butterworth}},\ }\bibfield  {title} {\bibinfo {title} {On the {{Theory}} of
  {{Filter Amplifiers}}},\ }\href@noop {} {\bibfield  {journal} {\bibinfo
  {journal} {Experimental Wireless \& the Wireless Engineer}\ }\textbf
  {\bibinfo {volume} {7}},\ \bibinfo {pages} {536} (\bibinfo {year}
  {1930})}\BibitemShut {NoStop}%
\bibitem [{\citenamefont {Franca}\ \emph {et~al.}(2023)\citenamefont {Franca},
  \citenamefont {Seidemann}, \citenamefont {Hassler}, \citenamefont {van~den
  Brink},\ and\ \citenamefont {Fulga}}]{Zenodo}%
  \BibitemOpen
  \bibfield  {author} {\bibinfo {author} {\bibfnamefont {S.}~\bibnamefont
  {Franca}}, \bibinfo {author} {\bibfnamefont {T.}~\bibnamefont {Seidemann}},
  \bibinfo {author} {\bibfnamefont {F.}~\bibnamefont {Hassler}}, \bibinfo
  {author} {\bibfnamefont {J.}~\bibnamefont {van~den Brink}},\ and\ \bibinfo
  {author} {\bibfnamefont {I.~C.}\ \bibnamefont {Fulga}},\ }\bibfield  {title}
  {\bibinfo {title} {Two-point spectroscopy of {F}ibonacci topoelectrical
  circuits},\ }\bibfield  {journal} {\bibinfo  {journal} {Zenodo}\ }\href
  {https://doi.org/10.5281/zenodo.8386622} {10.5281/zenodo.8386622} (\bibinfo
  {year} {2023})\BibitemShut {NoStop}%
\bibitem [{\citenamefont {Hua}\ and\ \citenamefont {Sarkar}(1990)}]{Hua1990}%
  \BibitemOpen
  \bibfield  {author} {\bibinfo {author} {\bibfnamefont {Y.}~\bibnamefont
  {Hua}}\ and\ \bibinfo {author} {\bibfnamefont {T.}~\bibnamefont {Sarkar}},\
  }\bibfield  {title} {\bibinfo {title} {Matrix pencil method for estimating
  parameters of exponentially damped/undamped sinusoids in noise},\ }\href
  {https://doi.org/10.1109/29.56027} {\bibfield  {journal} {\bibinfo  {journal}
  {IEEE Transactions on Acoustics, Speech, and Signal Processing}\ }\textbf
  {\bibinfo {volume} {38}},\ \bibinfo {pages} {814} (\bibinfo {year}
  {1990})}\BibitemShut {NoStop}%
\bibitem [{\citenamefont {Vanhamme}\ \emph {et~al.}(2001)\citenamefont
  {Vanhamme}, \citenamefont {Sundin}, \citenamefont {Hecke},\ and\
  \citenamefont {Huffel}}]{Vanhamme2001}%
  \BibitemOpen
  \bibfield  {author} {\bibinfo {author} {\bibfnamefont {L.}~\bibnamefont
  {Vanhamme}}, \bibinfo {author} {\bibfnamefont {T.}~\bibnamefont {Sundin}},
  \bibinfo {author} {\bibfnamefont {P.~V.}\ \bibnamefont {Hecke}},\ and\
  \bibinfo {author} {\bibfnamefont {S.~V.}\ \bibnamefont {Huffel}},\ }\bibfield
   {title} {\bibinfo {title} {{MR} spectroscopy quantitation: a review of
  time-domain methods},\ }\href
  {https://doi.org/https://doi.org/10.1002/nbm.695} {\bibfield  {journal}
  {\bibinfo  {journal} {NMR in Biomedicine}\ }\textbf {\bibinfo {volume}
  {14}},\ \bibinfo {pages} {233} (\bibinfo {year} {2001})}\BibitemShut
  {NoStop}%
\bibitem [{\citenamefont {Zieliński}\ and\ \citenamefont
  {Duda}(2011)}]{Zieliski2011}%
  \BibitemOpen
  \bibfield  {author} {\bibinfo {author} {\bibfnamefont {T.~P.}\ \bibnamefont
  {Zieliński}}\ and\ \bibinfo {author} {\bibfnamefont {K.}~\bibnamefont
  {Duda}},\ }\bibfield  {title} {\bibinfo {title} {Frequency and damping
  estimation methods - an overview},\ }\href
  {https://yadda.icm.edu.pl/baztech/element/bwmeta1.element.baztech-article-BSW1-0087-0001}
  {\bibfield  {journal} {\bibinfo  {journal} {Metrology and Measurement
  Systems}\ }\textbf {\bibinfo {volume} {18}},\ \bibinfo {pages} {505}
  (\bibinfo {year} {2011})}\BibitemShut {NoStop}%
\end{thebibliography}%


\begin{thebibliography}{5}%
\makeatletter
\providecommand \@ifxundefined [1]{%
 \@ifx{#1\undefined}
}%
\providecommand \@ifnum [1]{%
 \ifnum #1\expandafter \@firstoftwo
 \else \expandafter \@secondoftwo
 \fi
}%
\providecommand \@ifx [1]{%
 \ifx #1\expandafter \@firstoftwo
 \else \expandafter \@secondoftwo
 \fi
}%
\providecommand \natexlab [1]{#1}%
\providecommand \enquote  [1]{``#1''}%
\providecommand \bibnamefont  [1]{#1}%
\providecommand \bibfnamefont [1]{#1}%
\providecommand \citenamefont [1]{#1}%
\providecommand \href@noop [0]{\@secondoftwo}%
\providecommand \href [0]{\begingroup \@sanitize@url \@href}%
\providecommand \@href[1]{\@@startlink{#1}\@@href}%
\providecommand \@@href[1]{\endgroup#1\@@endlink}%
\providecommand \@sanitize@url [0]{\catcode `\\12\catcode `\$12\catcode
  `\&12\catcode `\#12\catcode `\^12\catcode `\_12\catcode `\%12\relax}%
\providecommand \@@startlink[1]{}%
\providecommand \@@endlink[0]{}%
\providecommand \url  [0]{\begingroup\@sanitize@url \@url }%
\providecommand \@url [1]{\endgroup\@href {#1}{\urlprefix }}%
\providecommand \urlprefix  [0]{URL }%
\providecommand \Eprint [0]{\href }%
\providecommand \doibase [0]{https://doi.org/}%
\providecommand \selectlanguage [0]{\@gobble}%
\providecommand \bibinfo  [0]{\@secondoftwo}%
\providecommand \bibfield  [0]{\@secondoftwo}%
\providecommand \translation [1]{[#1]}%
\providecommand \BibitemOpen [0]{}%
\providecommand \bibitemStop [0]{}%
\providecommand \bibitemNoStop [0]{.\EOS\space}%
\providecommand \EOS [0]{\spacefactor3000\relax}%
\providecommand \BibitemShut  [1]{\csname bibitem#1\endcsname}%
\let\auto@bib@innerbib\@empty
\bibitem [{\citenamefont {Hayt}\ and\ \citenamefont
  {Buck}(2006)}]{hayt2006engineering}%
  \BibitemOpen
  \bibfield  {author} {\bibinfo {author} {\bibfnamefont {W.}~\bibnamefont
  {Hayt}}\ and\ \bibinfo {author} {\bibfnamefont {J.}~\bibnamefont {Buck}},\
  }\href {https://books.google.fr/books?id=XwNJPgAACAAJ} {\emph {\bibinfo
  {title} {Engineering Electromagnetics}}}\ (\bibinfo  {publisher} {McGraw-Hill
  Higher Education},\ \bibinfo {year} {2006})\BibitemShut {NoStop}%
\bibitem [{\citenamefont {Franca}\ \emph {et~al.}(2023)\citenamefont {Franca},
  \citenamefont {Seidemann}, \citenamefont {Hassler}, \citenamefont {van~den
  Brink},\ and\ \citenamefont {Fulga}}]{zenodo}%
  \BibitemOpen
  \bibfield  {author} {\bibinfo {author} {\bibfnamefont {S.}~\bibnamefont
  {Franca}}, \bibinfo {author} {\bibfnamefont {T.}~\bibnamefont {Seidemann}},
  \bibinfo {author} {\bibfnamefont {F.}~\bibnamefont {Hassler}}, \bibinfo
  {author} {\bibfnamefont {J.}~\bibnamefont {van~den Brink}},\ and\ \bibinfo
  {author} {\bibfnamefont {I.~C.}\ \bibnamefont {Fulga}},\ }\bibfield  {title}
  {\bibinfo {title} {Two-point spectroscopy of {F}ibonacci topoelectrical
  circuits},\ }\bibfield  {journal} {\bibinfo  {journal} {Zenodo}\ }\href
  {https://doi.org/10.5281/zenodo.8386622} {10.5281/zenodo.8386622} (\bibinfo
  {year} {2023})\BibitemShut {NoStop}%
\bibitem [{\citenamefont {Hua}\ and\ \citenamefont {Sarkar}(1990)}]{Hua1990}%
  \BibitemOpen
  \bibfield  {author} {\bibinfo {author} {\bibfnamefont {Y.}~\bibnamefont
  {Hua}}\ and\ \bibinfo {author} {\bibfnamefont {T.}~\bibnamefont {Sarkar}},\
  }\bibfield  {title} {\bibinfo {title} {Matrix pencil method for estimating
  parameters of exponentially damped/undamped sinusoids in noise},\ }\href
  {https://doi.org/10.1109/29.56027} {\bibfield  {journal} {\bibinfo  {journal}
  {IEEE Transactions on Acoustics, Speech, and Signal Processing}\ }\textbf
  {\bibinfo {volume} {38}},\ \bibinfo {pages} {814} (\bibinfo {year}
  {1990})}\BibitemShut {NoStop}%
\bibitem [{\citenamefont {Hwee}(2013)}]{Hwee2013}%
  \BibitemOpen
  \bibfield  {author} {\bibinfo {author} {\bibfnamefont {C.~M.}\ \bibnamefont
  {Hwee}},\ }\emph {\bibinfo {title} {Matrix Pencil method as a signal
  processing technique performance and application on power systems signals}},\
  \href@noop {} {Ph.D. thesis},\ \bibinfo  {school} {National University of
  Singapore} (\bibinfo {year} {2013})\BibitemShut {NoStop}%
\bibitem [{\citenamefont {Assiimwe}\ \emph {et~al.}(2018)\citenamefont
  {Assiimwe}, \citenamefont {Mwangi},\ and\ \citenamefont
  {Konditi}}]{Assiimwe2018}%
  \BibitemOpen
  \bibfield  {author} {\bibinfo {author} {\bibfnamefont {E.}~\bibnamefont
  {Assiimwe}}, \bibinfo {author} {\bibfnamefont {E.}~\bibnamefont {Mwangi}},\
  and\ \bibinfo {author} {\bibfnamefont {D.~B.~O.}\ \bibnamefont {Konditi}},\
  }\bibfield  {title} {\bibinfo {title} {A matrix pencil method for the
  efficient computation of direction of arrival estimation for weakly
  correlated signals using uniform linear array in a low {SNR} regime},\ }\href
  {https://www.ripublication.com/irph/ijert18/ijertv11n9_01.pdf} {\bibfield
  {journal} {\bibinfo  {journal} {International Journal of Engineering Research
  and Technology}\ }\textbf {\bibinfo {volume} {11}},\ \bibinfo {pages} {1347}
  (\bibinfo {year} {2018})}\BibitemShut {NoStop}%
\end{thebibliography}%

\end{document}


\title{Supplemental Material to: \\ Two-point spectroscopy of Fibonacci topoelectrical circuits}

\author{Selma Franca}
\affiliation{Institute for Theoretical Solid State Physics, IFW Dresden and W\"urzburg-Dresden Cluster of Excellence ct.qmat, Helmholtzstr. 20, 01069 Dresden, Germany}
\affiliation{Institut Neel, CNRS, Grenoble, France}

\author{Torsten Seidemann}
\affiliation{Institute for Theoretical Solid State Physics, IFW Dresden and W\"urzburg-Dresden Cluster of Excellence ct.qmat, Helmholtzstr. 20, 01069 Dresden, Germany}

\author{Fabian Hassler}
\affiliation{JARA-Institute for Quantum Information, RWTH Aachen University, 52056 Aachen, Germany}

\author{Jeroen van den Brink}
\affiliation{Institute for Theoretical Solid State Physics, IFW Dresden and W\"urzburg-Dresden Cluster of Excellence ct.qmat, Helmholtzstr. 20, 01069 Dresden, Germany}
\affiliation{Institute  for  Theoretical  Physics,  TU  Dresden,  01069  Dresden,  Germany}

\author{Ion Cosma Fulga}
\affiliation{Institute for Theoretical Solid State Physics, IFW Dresden and W\"urzburg-Dresden Cluster of Excellence ct.qmat, Helmholtzstr. 20, 01069 Dresden, Germany}

\date{\today}
\begin{abstract}
In this Supplemental Material, we show details of the experimental realization before discussing how realistic capacitors and inductors affect the two point impedance. 
We proceed with discussing the matrix pencil method, before showing reconstructed spectra for the system under periodic boundary conditions.
\end{abstract}
\maketitle

\section{Details of the experimental realization}

\begin{figure*}[t]
\includegraphics[width=1\textwidth]{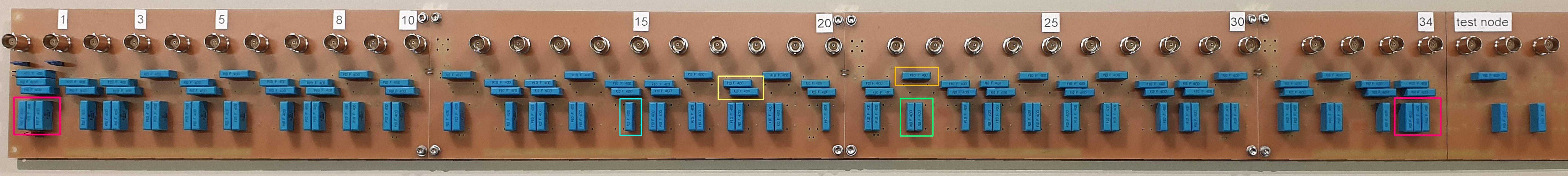}
\caption{A photograph of the Fibonacci topoelectrical chain. 
Yellow and orange boxes indicate capacitances $C_A=50 \rm nF$ and $C_B=100 \rm nF$, respectively.  
The capacitance $C_A= 50 \rm nF$ is realized by connecting, in series, two capacitors of capacitance $C_B$. 
Green and light blue boxes represent capacitances $C_A$ and $C_B$, respectively, used for grounding of the nodes in the bulk. 
In presence of open boundary conditions, two end nodes are grounded using capacitances $C_A+C_B=150 \rm nF$ represented with red boxes.
 }
\label{fig:figsys}
\end{figure*}

The characteristic function $\chi_n (\phi=\pi)$ defined in Eq.~(2) of the main text produces the following configuration of hopping amplitudes 
\begin{equation*}
\rm{AABABAABABAABAABABAABAABABAABABAA}
\end{equation*}
for a $N=34$ sites long chain. 
Here, letters $A$ and $B$ represent hoppings $t_A$ and $t_B$, respectively.
The corresponding Fibonacci topelectrical circuit is shown in Fig.~\ref{fig:figsys}.
As explained in the main text, hopping amplitudes $t_A$ and $t_B$ are emulated with capacitors of effective capacitances $50 \rm nF$ and $100 \rm nF$, respectively.
The electrical circuit, however, was built using a single type of capacitors with nominal capacitances $100 \rm nF$ bought from KEMET manufacturer (product number F461BC104G400C).
To obtain the capacitance $C_A= 50 \rm nF$, we connect in series two capacitors of capacitance $C_B = 100 \rm nF$.
In Fig.~\ref{fig:figsys}, yellow and orange boxes indicate circuit elements with capacitances $C_A$ and $C_B$, respectively.

Besides using capacitors to emulate hoppings of the Fibonacci chain, we use them to ground of the nodes. 
The circuit is constructed such that the grounding capacitances $\tilde{C}_n$ fulfill equality $\tilde{C}_n+C_{n-1}+C_{n} = 2C_A+C_B$.
Inside the bulk of the circuit, where $C_{n-1}, C_n \in \{C_A, C_B\}$, the grounding capacitances take values $\tilde{C}_n \in \{C_A, C_B\}$ depending on the local environment, see also Fig.~1(a) of the main text.
We indicate these two kinds of grounding capacitances, $C_A$ and $C_B$ with green and blue boxes in Fig.~\ref{fig:figsys}, respectively.
However, for a finite system the grounding capacitances at the ends are $\tilde{C}_1=\tilde{C}_{34}= C_A+C_B$ because either $C_{n-1}$ or $C_n$ equal zero.
We indicate these two grounding capacitances with red boxes in Fig.~\ref{fig:figsys}.
Therefore, to build a circuit of $34$ sites with open boundary conditions we need $116$ capacitors in total. 
To obtain a chain under periodic boundary conditions, we relate the first and the last node via a capacitor of capacitance $C_A$.
This changes the grounding capacitances of these nodes to $\tilde{C}_1=\tilde{C}_{34}= C_B$, implying that we need $114$ capacitors to build a circuit with periodic boundary conditions.
Capacitors have the average capacitance $\bar{C} = 100.22 \rm nF$ with the standard deviation $\sigma_C = 0.56 \rm nF$. 
For a randomly chosen capacitor used in the circuit, we find that its capacitance and capacitive reactance are stable to changes in alternative current frequency, see Table~\ref{tab:table1}.
For this reason, we denote capacitive reactance as $R_C^{\rm dc}$ and take its value to be $R_C^{\rm dc}  = 25 \rm m\Omega$ (nominal value given by the manufacturer).
\begin{table}[h!]
    \centering
\begin{tabular}{ c c c c c c}
\hline
 $f$(kHz) & 50 & 100 & 150 &200 & 250  \\ [0.5ex] 
 \hline
$C$ (nF)  & 100.046 & 100.08  & 100.15 & 99.7 & 99.75  \\ 
 $R$ $(\rm m\Omega)$  & 11 & 27 & 31 &32 &22 
\end{tabular}
\label{tab:table1}
\caption{Measured values of capacitance and capacitive reactance for a single capacitor of KEMET manufacturer.}
\end{table} 

\begin{figure}[h!]
\includegraphics[width=1\columnwidth]{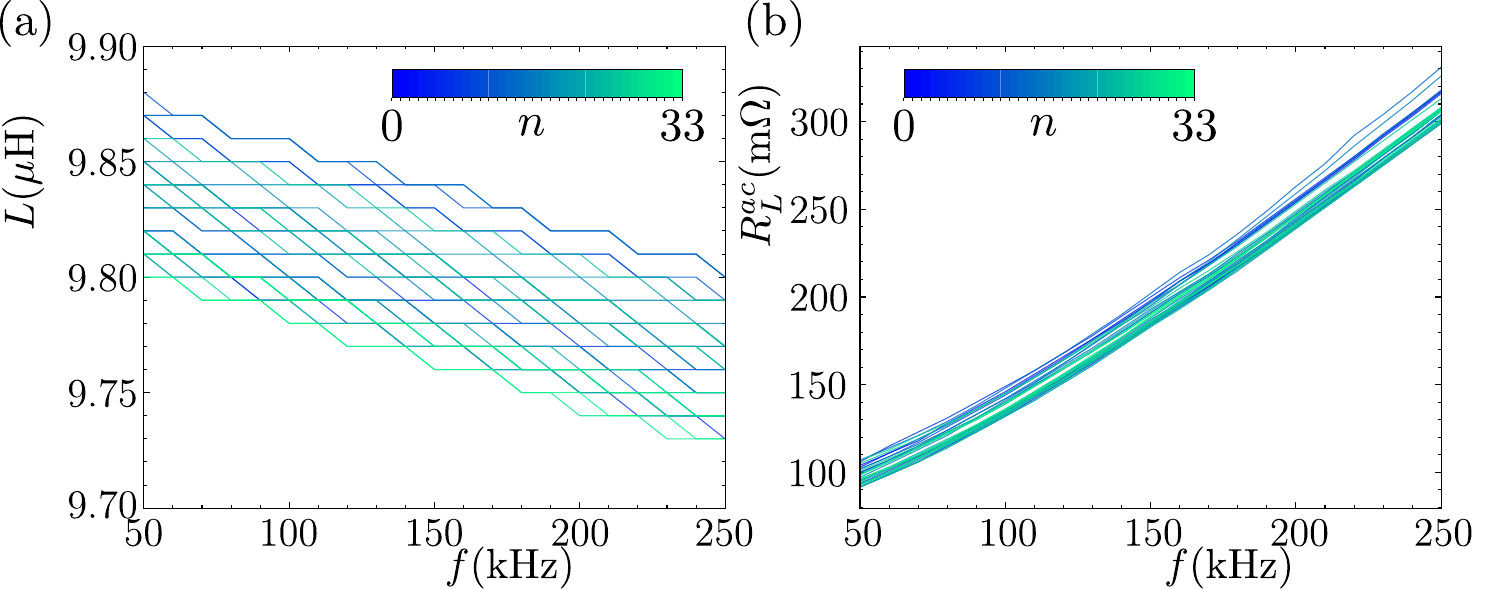}
\caption{(a) Dependence of inductance on frequency for all inductors used in the circuit. 
(b) Corresponding inductive reactances as a function of frequency. }
\label{fig:figsm0}
\end{figure}

Inductors used in grounding of the nodes are WURTH Elektronik LHMI SMT Power Inductor 7030 (order code 74437346100) with nominal resistance of $10 \rm \mu H$ at $f =100 $ kHz. 
Since our measurements require sweeps over frequency, we have measured how the inductance of every inductor changes with frequency. 
Results are shown in Fig.~\ref{fig:figsm0}(a), and we see that the inductance reduces with frequency due to the skin effect~\cite{hayt2006engineering}.
Namely, the average inductance at $f=50$ kHz is $\bar{L}(50 \rm kHz) = 9.84 \mu H$ with the standard deviation of $\sigma_L (50 \rm kHz) = 0.22 nH$, 
while at $f= 250 \rm kHz$, $\bar{L}(250 \rm kHz) = 9.76 \mu H$ and the standard deviation equals $\sigma_L (250 \rm kHz) = 0.23 nH$.
Moreover, the resistance upon passing of the direct current is $R_C^{dc} = 85 \rm m\Omega$ according to the manufacturer.
Lastly, we have measured the inductive reactance $R_L^{ac}$ of all capacitors as a function of frequency.
From Fig.~\ref{fig:figsm0}(b), we see it increases significantly with frequency due to the aforementioned skin effect.
More precisely, the average resistance at $f=50\rm kHz$ is $\bar{ R}_L^{ac} (50 \rm kHz) = 97.4 \rm m\Omega$ with the standard deviation of $\sigma_{R_L}  (50 \rm kHz)= 4.67 \rm m \Omega$.
For $f= 250\rm kHz$, it equals $\bar{R}_C^{ac} (250 \rm kHz) = 308 \rm m\Omega$ with the standard deviation being $\sigma_{R_L} (250 \rm kHz) = 7.6 \rm m \Omega$.

\section{Simulations of realistic impedance}

In ideal case, the two-point impedance between nodes $a$ and $b$ can be calculated from eigenvalues $Y_n$ and eigenvectors $v_n$ of the admittance matrix $Y(f)$ as
\begin{equation}
Z_{a,b} (f) = \sum_n \frac{|v_{n,a}-v_{n,b}|^2}{Y_n},
\end{equation}
see Eq.~(4) of the main text.
Here, the admittance matrix is defined as $Y(f) = \tilde{g}(f) \mathbb{I} - 2\pi j H$, with $\tilde{g}(f) = 2\pi j f (2C_A+C_B) +1/(2\pi j f L)$.
In order to capture effects of nonzero resistances of capacitors and inductors, it is advantageous to use the \textsc{LTspice} software for simulations of the topoelectrical Fibonacci chain as it allows for straightforward implementation of these experimental imperfections. 
On Zenodo~\cite{zenodo}, we share our netlist files of the Fibonacci topoelectrical circuit.

\begin{figure}[tb!]
\includegraphics[width=0.9\columnwidth]{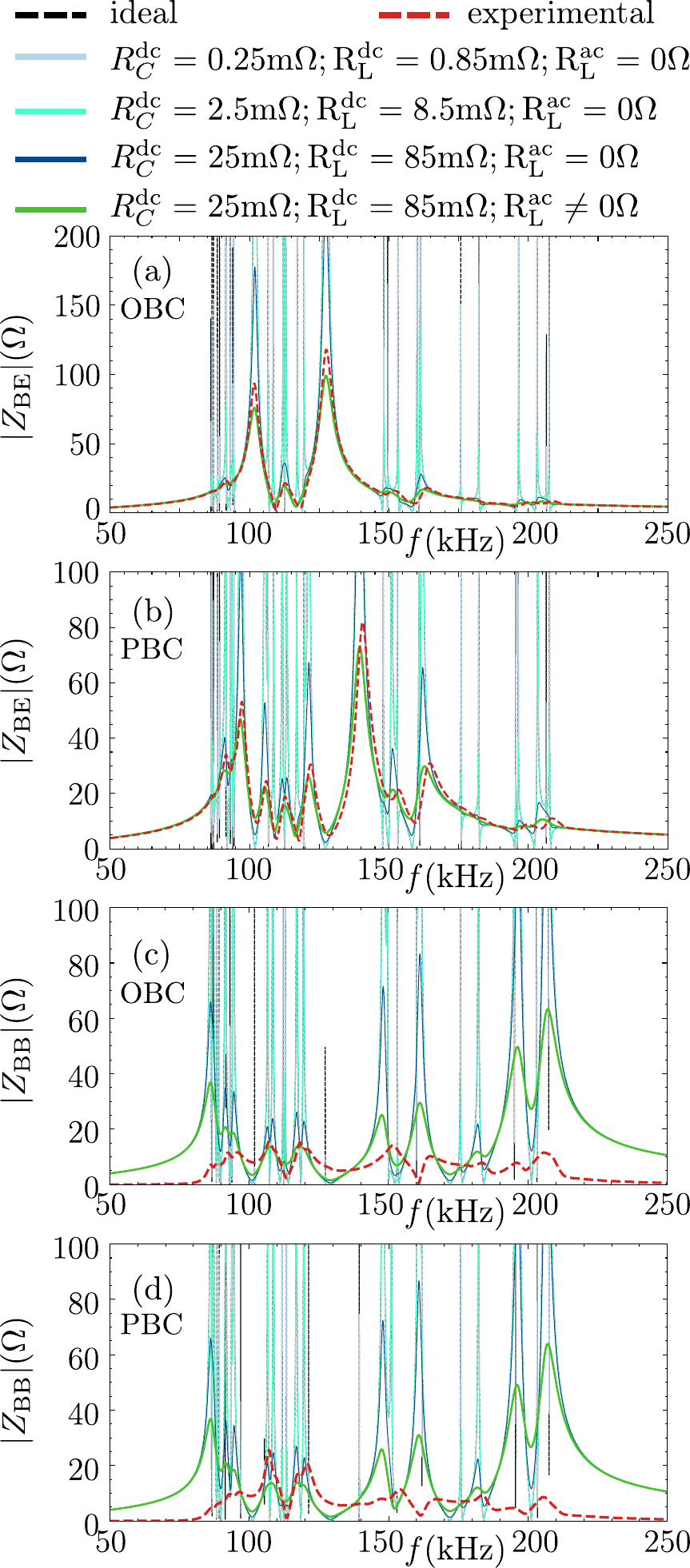}
\caption{The comparison between simulated and experimentally measured response function $Z_{a,b}(f)$. 
Panels (a)/(c) and (b)/(d) concern the BE/BB configuration of voltage probes under open and periodic boundary conditions, respectively.
Here, OBC denotes open boundary conditions and PBC periodic boundary conditions.
 }
\label{fig:figsm1}
\end{figure}

In the following, we study how realistic resistances of capacitors and inductors affect the impedance response.
We focus first of the BE configuration of voltage probes that involves nodes $a= 1$ and $b= 15$.
In Fig.~\ref{fig:figsm1}(a) are plotted $|Z_{\rm BE}|(f)$ for different strengths of resistances $R_C^{\rm dc}, R_L^{\rm dc}$ and $R_L^{\rm ac}$ in case of open boundary conditions.
Dashed black lines represent the voltage response of an ideal circuit, with all resistances equal zero.
We see that such response is a linear combination of the-delta-peaks placed at resonance frequencies $f_n$.
We first study effects of resistances $R_C^{\rm dc}$ and $R_L^{\rm dc}$ by gradually increasing their strength towards the experimental values.
Resistance strengths $R_C^{\rm dc}=25 \rm m\Omega/100$ and $R_L^{\rm dc}=85 \rm m\Omega/100$ reduce the height of some delta-peaks but preserve their width such that they remain well separated from each other in frequency space.
However, when the resistance strengths reach $1/10$th of the experimental values, i.e., $R_C^{\rm dc}=25 \rm m\Omega/10$ and $R_L^{\rm dc}=85 \rm m\Omega/10$, the-delta-peaks broaden and can even merge with their neighbors if this broadening is larger than the distance between two peaks.
Therefore, these resistance strengths yield a smaller number of Lorentzian features in the response function compared to the number of theoretical eigenvalues. 
For experimentally relevant strengths of $R_C^{\rm dc}$ and $R_L^{\rm dc}$, the-delta-peaks of the ideal response are replaced with broadly peaked features that resemble in position and number to the peaks of the experimentally measured two-point impedance. 
Simulation can be further improved by adding nonzero $R_L^{\rm ac}$, and we see that the resulting simulation (green curve in Fig.~\ref{fig:figsm1}(a)) captures well the experimental result.

In Fig.~\ref{fig:figsm1}(b) are plotted simulations and experiment measurement of $|Z_{\rm BE}|(f)$ for the topoelectrical Fibonacci chain under periodic boundary conditions.
As before, the simulation in \textsc{LTSpice} that takes into account realistic values of three different resistances $R_C^{\rm dc}, R_L^{\rm dc}$ and $R_L^{\rm ac}$ (green curve) capture well the experimental result.
Note that all these response functions do not have pronounced features at the frequencies corresponding to edge modes for which we observe two strongest peaks in Fig.~\ref{fig:figsm1}(a), confirming that the circuit is under the periodic boundary conditions. 

In Figs.~\ref{fig:figsm1}(c) and Fig.~\ref{fig:figsm1}(d) are shown results for the two-point impedance between nodes $a = 10$ and $b=24$  in case of open and periodic boundary conditions, respectively.
We observe first that the amplitude of the measured response function does not change significantly in the frequency range.
This indicates that none of the eigenstates dominate the response, resulting in the two-point impedance of the BB probe having more features compared to the response measured by the BE probe.
For open boundary conditions, we observe that the simulation that takes into account all the resistances captures well the amplitude and peaks of the measured impedance for frequencies  $100 \rm kHz <$ $ f  < 180 \rm kHz$, see Fig.~\ref{fig:figsm1}(c).
For $f \leq 100 \rm kHz$ and $f \geq 180 \rm kHz$, the simulation does not capture well the amplitude of the response function but recovers many peaks of the measured impedance. 
Similar conclusions can be reached for periodic boundary conditions, see Fig.~\ref{fig:figsm1}(d).
It is not surprising that the biggest difference between theory and simulation occurs for either small or large frequencies.
In the former case, the grouping of states in addition to noise might lead to some unforeseen effects that cannot be captured by the simulation. 
For larger frequencies however, we know that our simulations do not take into account all the frequency induced changes in the circuit elements.
For example, we ignore the changes in the inductance of inductors that were shown to occur in Fig.~\ref{fig:figsm0}(b).

\section{Matrix pencil method}

As explained in the main text, the matrix pencil method relies on fitting a time domain signal $Z_{a,b}^{(2)}(t)$ to the sum of $N$ damped complex exponentials.
As we measured the signal $Z_{a,b}(f)$ only in the frequency domain, to obtain the time domain signal $Z_{a,b}^{(2)}(t)$ we use a Fourier transform $\mathcal{F}$ as $Z_{a,b}^{(2)}(t) = \mathcal{F} [- \partial_f^2 Z_{a,b}]$. 
Performing the Fourier transformation requires knowing a zero frequency term of the signal $Z_{a,b} (f=0)$, as well as negative frequency terms that can be obtained with mapping $Z_{a,b} (-f) = Z^*_{a,b} (f)$. 
Since the circuit is made of capacitors and inductors that do not conduct direct current, we assume $Z_{a,b} (0) = 0$.
For the BB configuration of voltage probes, we have measured the signal in the frequency range $[10 \rm{Hz}, 350 kHz]$ with $10 \rm Hz$ increment.
In case of the BE configuration of voltage probes, the signal was measured with $10\rm Hz$ increment in the frequency range $ [50 \rm{kHz}, 250 kHz]$.
Since $Z(50\rm kHz) \neq 0$ for the BE configuration, assuming $Z(f<50\rm kHz) = 0$ would create a discontinuity in the $Z(f)$ signal that would reflect on the results of the matrix pencil method. 
For this reason, we assume that $\Re[Z_{a,b}(f)]$ and $\Im[Z_{a,b}(f)]$ grow linearly from zero to values $\Re[Z_{a,b}(50 \rm kHz)]$ and $\Im[Z_{a,b}(50 kHz)]$, respectively, in the frequency range $[0,50 \rm kHz)$ with an increment of $10\rm Hz$.
This choice of interpolation is retroactively verified upon confirming that the matrix pencil method on $Z_{BE}^{(2)}(t)$ does not find a resonance at $f=50 \rm kHz$.
For both voltage configurations, we study signal up till $f=250 \rm kHz$ such that its length is $N_s= 25000$.
This implies that the signal $Z_{a,b}^{(2)}(t)$ has length $2N_s+1 = 50001$.
To reduce the computational cost associated with such a huge number of entries for the matrix pencil method, we sample our data such that the frequency increment for the analyzed data becomes $50 \rm Hz$, i.e., every $5$th measured value is taken into account. 
This results in the discrete time domain signal of length $2N_s+1=10001$.

In the following, we explain the basics of the matrix pencil method.
We closely follow Refs.~\onlinecite{Hua1990, Hwee2013}.
For completeness, we rewrite the Eq.~(6) of the main text
\begin{equation}
Z^{(2)} (t) = \sum_{n=1}^N A^{\rm exp}_n  e^{j \phi^{\rm exp}_n} e^{(\alpha^{\rm exp}_n + 2\pi j f^{\rm exp}_n) t},
\end{equation}
with $A^{\rm exp}_n, \phi^{\rm exp}_n,\alpha^{\rm exp}_n,f^{\rm exp}_n$ representing amplitudes, phases, dampings and frequencies of sinusoids used to fit the measured data.
By setting $t = m T$, where $m = 0,..., 2 N_s$ denotes the sample number and $T$ the sampling period, a time-domain signal becomes
\begin{equation}
Z_{a,b}^{(2)}(mT) = Z_{a,b}^{(2)}(m) \approx \sum_{n = 1}^{N} A^{\rm exp}_n  e^{j \phi^{\rm exp}_n} z_n^m,
\end{equation}
where $z_n = e^{(\alpha^{\rm exp}_n + 2\pi j f^{\rm exp}_n) T}$. 
For a noiseless signal, poles $z_n$ can be found by solving the generalized eigenvalue problem 
\begin{equation}~\label{eq:matpenc}
(X_2 -z X_1) \Psi = 0,
\end{equation}
where Henkel matrices $X_1$ and $X_2$ are constructed from sampled data as
\begin{widetext}
\begin{equation}
X_1 = \begin{pmatrix}
Z_{a,b}^{(2)}(0) & Z_{a,b}^{(2)}(1) & \ldots & Z_{a,b}^{(2)}(M-1) \\
Z_{a,b}^{(2)}(1) & Z_{a,b}^{(2)}(2) & \ldots & Z_{a,b}^{(2)}(M) \\
\vdots & \vdots & \; & \vdots \\
Z_{a,b}^{(2)}(2N_s-M) &Z_{a,b}^{(2)}(2N_s-M+1) & \ldots & Z_{a,b}^{(2)}(2N_s-1)
\end{pmatrix}
\end{equation}
\end{widetext}
and 
\begin{widetext}
\begin{equation}
X_2 = \begin{pmatrix}
Z_{a,b}^{(2)}(1) & Z_{a,b}^{(2)}(2) & \ldots &Z_{a,b}^{(2)}(M) \\
Z_{a,b}^{(2)}(2) & Z_{a,b}^{(2)}(3) & \ldots & Z_{a,b}^{(2)}(M+1) \\
\vdots & \vdots & \; & \vdots \\
Z_{a,b}^{(2)}(2N_s+1-M) & Z_{a,b}^{(2)}(2N_s+2-M) & \ldots & Z_{a,b}^{(2)}(2N_s)
\end{pmatrix}.
\end{equation}
\end{widetext}
We call matrix $\mathbf{X} = X_1-zX_2$ the matrix pencil~\cite{Hwee2013}.
Matrices $X_1$ and $X_2$ are of dimensions $ (2N_s+1-L)\times M$, 
where $M$ is the pencil parameter that can take arbitrary values.
The matrix pencil method is least sensitive to noise for $M=(2N_s+1)/3$ and $M=2\times(2N_s+1)/3$~\cite{Hua1990, Hwee2013}.
Here, we choose $M= (2N_s+1)/3$.  

In case of a noisy signal such as ours, we need to discard noise before solving Eq.~\eqref{eq:matpenc}~\cite{Hua1990, Hwee2013}. 
This is done in the following way.
First, we construct matrix $X$ as
\begin{widetext}
\begin{equation}
X = \begin{pmatrix}
Z_{a,b}^{(2)}(0) & Z_{a,b}^{(2)}(1) & \ldots & Z_{a,b}^{(2)}(M) \\
Z_{a,b}^{(2)}(1) & Z_{a,b}^{(2)}(2) & \ldots & Z_{a,b}^{(2)}(M+1) \\
\vdots & \vdots & \; & \vdots \\
Z_{a,b}^{(2)}(2N_s-M) & Z_{a,b}^{(2)}(2N_s-M+1) & \ldots & Z_{a,b}^{(2)}(2N_s)
\end{pmatrix},
\end{equation}
\end{widetext}
and then we decompose it using the singular value decomposition
\begin{equation}\label{eq:svd}
X= U D V^{\dagger}.
\end{equation}
Here, unitary matrices $U$ and $V$ have $(2N_s+1-M) \times (2N_s+1-M)$ and $(M+1) \times (M+1)$ entries, respectively.
Matrix $D$ is of dimensions $(2N_s+1-M) \times (M+1)$ that has nonzero diagonal entries $\sigma_n$ that are non-negative real numbers sorted in non-increasing order ($\sigma_1 \geq \sigma_2 \geq \hdots$).
We can now split matrix $X$ into signal and noise subspace as~\cite{Assiimwe2018}
\begin{equation}
X = U^s D^s  (V^{s})^{\dagger}+ U^n D^n  (V^{n})^{\dagger},
\end{equation}
where the dimension of the signal subspace is determined by the number $N=34$ of damped exponentials.
This implies that matrix $D^s$ is constructed from $2N$ largest singular values $\sigma_n$, while matrices $U^s$ and $V^s$ are of dimensions $(2N_s+1-M) \times 2N$ and $2N \times (M+1)$, respectively.
From $X^s = U^s D^s (V^{s})^{\dagger}$, it is possible to construct matrices $X^s_1$ and $X^s_2$ by removing last and first columns, respectively.
The generalized eigenvalues of this matrix pencil $X^s_1 -z X^s_2$ can then be found by solving the following eigenvalue problem
\begin{equation}
(X^s_2)^{+}  X^s_1 \Psi = z \Psi,
\end{equation}
where $(X^s_2)^{+}$ is the Moore-Penrose inverse of matrix $X^s_2$ of rank $2N$. 
The Moore-Penrose inverse represents a generalization of matrix inverse for non-square matrices such as $X^s_2$, and is defined as $(X^s_2)^{+} = V_{s_1}^{2N} D_{s_1}^{2N} (U_{s_1}^{2N})^{ \dagger}$ for rank $2N$. 
Here, $V_{s_1}, D_{s_1}, U_{s_1}$ are matrices corresponding to the singular value decomposition of matrix $X_{s_1}$, see also Eq.~\eqref{eq:svd}.
Note that the diagonal entries of $D_{s_1}$ are arranged in non-increasing order, such that $D_{s_1}^{2N}$ contains only $2N$ largest singular values $\sigma_n$ and $V_{s_1}^{2N}$ and $U_{s_1}^{2N}$ the corresponding singular vectors.
The frequencies $f_n^{\rm exp}$ and corresponding damping factors $\alpha_n^{\rm exp}$ can be calculated from $\Im[\log{z_n}]$ and $\Re[\log{z_n}]$, respectively, while amplitudes $A_n$ and phases $\phi_n$ are found by solving the least squared method~\cite{Hwee2013}.

\section{Spectrum reconstruction for a topoelectrical Fibonacci chain under periodic boundary conditions}

\begin{figure*}[tb]
\includegraphics[width=1\textwidth]{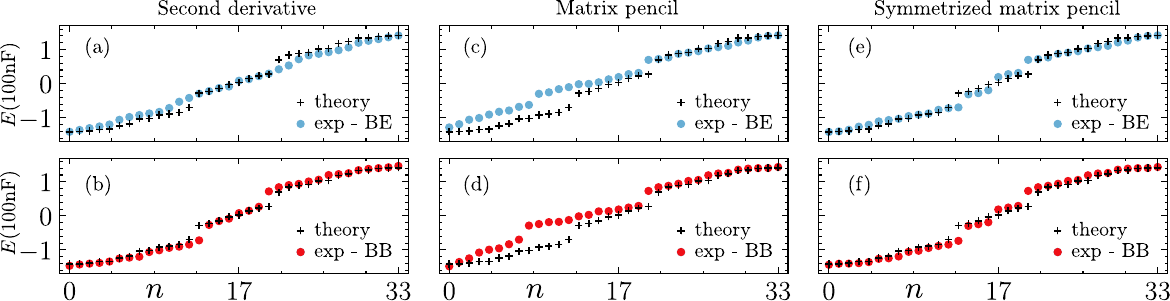}
\caption{
Comparison between theoretical and experimental spectra obtained using different methods of recovery. 
Panels (a) and (b) are obtained using maxima of $-\partial_f^2 |Z_{\rm BE}|$ and $-\partial_f^2 |Z_{\rm BB}|$.
Panels (c) and (d) are produced using $34$ resonant frequencies that were detected using the matrix pencil method. 
Lastly, panels (e) and (f) are obtained by mirroring, with respect to $E=0$, all $17$ positive eigenvalues from panels (c) and (d), respectively. 
 }
\label{fig:fig3}
\end{figure*}

The experimental eigenvalues in case of PBCs are shown in Fig.~\ref{fig:fig3} for different methods of reconstruction.
Our first method, as explained in the main text, relies on searching for $17$ most prominent peaks of $-\partial_f^2 |Z_{a,b}|$ in the frequency range $(f_0, 250 \rm kHz)$ that corresponds to the positive part of the spectrum.
After all positive eigenvalues are identified, the total spectrum can be recovered using the property of spectral symmetry with respect to $E=0$. 
To filter the noise, for both data sets we employ the $4$-th order Butterworth filter with the cutoff frequency $f_c = 0.01 f_{N_q}$, i.e., using the same parameters as in the main text.

The results are shown in Fig.~\ref{fig:fig3}(a) and (b) for the BE and BB configurations of voltage probes, respectively.
In the case of the BE configuration, we observe that this method yields a gapless spectrum and hence does not capture well two major gaps in the theoretical spectrum.
In particular, the eigenvalues forming the middle band are captured well by experiment as well as the topmost eigenvalues forming the upper band.
Due to the $\mathcal{C}$ symmetry, this implies that the lowest eigenvalues of the lower band are also measured well in experiment.
However, as $|E|$ is reduced, the accuracy of the method declines for the upper and lower bands. 
In total, the mean absolute error is $\delta_{\rm BE}^{\rm avg} = 11.32 \rm nF$ and the related median error equals $\delta_{\rm BE}^{\rm m} = 8.78 \rm nF$.
The accuracy of this method increases for the BB probe, as we observe the existence of three bands separated by two gaps.
Most of the eigenvalues are well captured by the experiment, except two modes of the middle band that are wrongly attributed to the upper and lower bands (one mode per band).
The mean absolute error for this measurement equals $\delta_{\rm BB}^{\rm avg} = 7.45 \rm nF$ with a much smaller median error $\delta_{\rm BB}^{\rm m} = 4.41 \rm nF$ indicating that these outliers strongly affect the error statistics.

The matrix pencil method allows us to measure all $34$ eigenvalues.
To eliminate noise from $-\partial_f^2 Z_{a,b}$, we use again the $4$-th order Butterworth filter with the cutoff frequency $f_c= 0.03f_{N_q}$ for the BB probe and $f_c = 0.01 f_{N_q}$ for the BE probe.
The results are shown in Fig.~\ref{fig:fig3}(c) and (d) for the BE and BB configurations of voltage probes, respectively.
As was the case for the circuit under open boundary conditions (see Fig.~3 of the main text), the results become less accurate as the energy is decreased, with the upper band being captured the best by the measurement for both setups.
For both setups, this method attributes an additional eigenvalue to the upper band at the expense of the middle band while simultaneously attributing more eigenvalues to the middle band at the expense of the lower band. 
In total, the mean absolute errors are  $\delta_{\rm BE}^{\rm avg} = 23.01 \rm nF$ and $\delta_{\rm BB}^{\rm avg} =22.56\rm nF$ with median errors $\delta_{\rm BE}^{\rm m} = 15.96 \rm nF$ and $\delta_{\rm BB}^{\rm m} = 15.02 \rm nF$.

To improve our results obtained using the matrix pencil method, we use the fact the spectrum should be symmetric with respect to $E=0$.
Since the former method finds $17$ positive eigenvalues for both voltage probe configurations, the total spectrum is obtained by combining these eigenvalues with their negative counterparts.
The results are shown in Fig.~\ref{fig:fig3}(e) and (f) for the BE and BB configurations of voltage probes, respectively.
Applying this symmetry reduces the errors of measurement to $\delta_{\rm BE}^{\rm avg} = 8.75 \rm nF$ and $\delta_{\rm BB}^{\rm avg} =8.79 \rm nF$, while the median errors become $\delta_{\rm BE}^{\rm m} = 7.41 \rm nF$ and $\delta_{\rm BB}^{\rm m} =5.48 \rm nF$.
Therefore, the matrix pencil method in combination with chiral symmetry constraint produces a more accurate measurement for the BE probe compared to our first method based on searching for the peaks of $-\partial_f^2 |Z_{\rm BE}|$.
This contrasts the results obtained for the BB configuration of probes, as our first method outperforms both matrix pencil methods.
\newpage 
\bibliography{NHr}